\documentclass[aps,twocolumn,amsmath,letterpaper]{revtex4}
\usepackage{amssymb}
\usepackage{graphicx}
\usepackage{array}
\usepackage{hhline}
\usepackage{longtable}

\renewcommand{\thetable}{\Roman{table}} \thetable
\newcommand{\<}[1]{\hspace{-0.33333em}#1\hspace{-0.33333em}}
\newcommand{\rsp}[1]{\hspace{-0.15em}#1\hspace{-0.15em}}
\begin{document}

\title{Two Superconducting Phases in the $d=3$ Hubbard Model:\\ Phase Diagram and Specific Heat from Renormalization-Group Calculations}
\author{Michael Hinczewski and A. Nihat Berker}
\affiliation{Department of Physics, Istanbul Technical University,
Maslak 34469, Istanbul, Turkey,} \affiliation{Department of Physics,
Massachusetts Institute of Technology, Cambridge, Massachusetts
02139, U.S.A.,} \affiliation{Feza G\"ursey Research Institute,
T\"UBITAK - Bosphorus University, \c{C}engelk\"oy 81220, Istanbul,
Turkey}
\begin{abstract}
The phase diagram of the $d=3$ Hubbard model is calculated as a
function of temperature and electron density $\langle n_i\rangle$,
in the full range of densities between 0 and 2 electrons per site,
using renormalization-group theory.  An antiferromagnetic phase
occurs at lower temperatures, at and near the half-filling density
of $\langle n_i\rangle = 1$.  The antiferromagnetic phase is
unstable to hole or electron doping of at most 15\%, yielding to two
distinct "$\tau$" phases:  for large coupling $U/t$, one such phase
occurs between 30-35\% hole or electron doping, and for small to
intermediate coupling $U/t$ another such phase occurs between
10-18\% doping.  Both $\tau$ phases are distinguished by non-zero
hole or electron hopping expectation values at all length scales.
Under further doping, the $\tau$ phases yield to hole- or
electron-rich disordered phases. We have calculated the specific
heat over the entire phase diagram. The low-temperature specific
heat of the weak-coupling $\tau$ phase shows a BCS-type exponential
decay, indicating a gap in the excitation spectrum, and a cusp
singularity at the phase boundary. The strong-coupling $\tau$ phase,
on the other hand, has characteristics of BEC-type
superconductivity, including a critical exponent $\alpha \approx
-1$, and an additional peak in the specific heat above the
transition temperature indicating pair formation.  In the limit of
large Coulomb repulsion, the phase diagram of the $tJ$ model is
recovered.

\end{abstract}
\maketitle
\def\s{\rule{0in}{0.28in}}

\section{Introduction}

\setlength{\LTcapwidth}{\columnwidth}

The Hubbard model \cite{Hubbard} is the simplest realistic (in that it
retains particulate dynamics) model of electronic conduction systems. This
model should constitute a fair description for many real solid-state physics
systems and a starting-point description for those systems with added
complexities such as quenched randomness, frustration, and/or spatial
anisotropy. The first query that comes to mind, in the study of either
experimental or model systems, is on the phase diagram, as a function of
physical parameters such as temperature and density. Nevertheless, until
recently \cite{Hubshort}, no estimates, let alone (be it approximate)
solutions, were ventured on the phase diagram of the Hubbard model at
dimensions greater than $d=1$, at temperatures greater than $T=0$, and densities
away from half-filling.

The first approach to a phase diagram problem, in the past before
the advent of renormalization-group theory \cite{Wilson}, had been
through a mean-field approximation.  However, such method is not
useful for the Hubbard model, since, where the characteristic
phenomena occur away from half-filling, the off-diagonal term in the
Hamiltonian plays a determining role, as we shall see below. There
is no ready way to deal with such a dominant quantum mechanical
effect using mean-field theory. On the other hand,
renormalization-group theory, which some time ago has excelled over
mean-field theory in phase diagram studies, is effective. Previous
renormalization-group calculations have concentrated on studying the
Hubbard model in lower dimensions, at zero temperature, or at
half-filling:  The zero-temperature (ground-state) properties were
successfully obtained in $d=1,2,3$.\cite{Hirsch, FourSpronk}  In
$d=1$ at half-filling, the thermodynamic properties were accurately
calculated for finite temperatures.\cite{VdzStella}  In cases where
comparison is possible due to the availability of exact results in
$d=1$, the renormalization-group results have proven to be very
accurate, coming to within about 1\% of the exact
results.\cite{FourSpronk,VdzStella}  In $d=2$ at half filling, it
was found that no phase transition occurs as a function of
temperature.\cite{Vdz,Cannas1}  This result was later extended to
other fillings in $d=2$ \cite{Cannas2} and confirmed by quantum
Monte Carlo calculations \cite{Cosentini}. In $d=3$ at half filling,
an antiferromagnetic phase transition as a function of temperature
was obtained.\cite{Cannas1}  One calculation done in $d=3$ at finite
temperature and arbitrary chemical potential \cite{Cannas2} did not
obtain the "$\tau$" phase reported below and in Ref.\cite{Hubshort}.

The physics of the Hubbard model in the limit of large Coulomb
repulsion is believed to be described by the $tJ$
model~\cite{Anderson,BZA}. Application of renormalization-group
theory to the entire density range of the $tJ$ model at finite
temperatures in $d=3$ has yielded
\cite{FalicovBerker,FalicovBerkerT}, between 30-40\% vacancies from
$\langle n_i\rangle =1$, a novel (dubbed "$\tau$") phase in which
the electron hopping strength in the Hamiltonian renormalizes to
infinity under repeated scale changes, while the system remains
partially filled. The calculated topology of the phase diagram,
including near the $\tau$ phase a first-order phase transition that
is very narrow (less than 2\% jump in the electron density) and an
antiferromagnetic phase that is unstable to at most 10\% vacancies
from $\langle n_i\rangle =1$, is indeed reminiscent of experimental
phase diagram determinations with lanthanide oxides \cite{ChouJohn}.

While the studies
above~\cite{Hirsch,FourSpronk,VdzStella,Vdz,Cannas1,Cannas2,FalicovBerker,FalicovBerkerT}
have used position-space renormalization-group approaches, there has
recently been a revival of interest in Wilson perturbative
renormalization-group methods applied to correlated fermion
problems.  These methods have long been known to be successful for
one-dimensional systems~\cite{Solyom,Voit} and, in the last few
years, for the $d=2$ Hubbard model, they have yielded
antiferromagnetic instabilities near half-filling and
superconducting instabilities at smaller
densities~\cite{ZanchiSchulzA,ZanchiSchulzB,HalbothMetznerA,HalbothMetznerB,HonerkampA,HonerkampB}.
Because of the perturbative nature of these treatments, their
predictions are strictly valid only in the case of weak coupling.
The position-space renormalization-group method presented in this
paper appears to work over the entire range of coupling strengths,
as seen below, and yields definite phase diagrams and thermodynamic
functions.

In fact, our approach makes an interesting prediction for the
evolution of the Hubbard phase diagram as coupling is increased. We
find two distinct $\tau$ phases, one occurring at small to
intermediate coupling and the other, inclusive of the $tJ$ model
$\tau$ phase, occurring at strong coupling.  From an analysis of
their specific heat behaviors, we find that the two $\tau$ phases
respectively have characteristic properties of a weakly-coupled
BCS-type and a strongly-coupled BEC-type superconducting phase.
Since high-$T_c$ materials share aspects of both limits, and are
thought to lie in some intermediate coupling range~\cite{Junod}, our
prediction for the Hubbard phase diagram may be directly relevant to
the physics of high-$T_c$ superconductors.

\section{The Hubbard Model}

The Hubbard model is defined by the Hamiltonian

\begin{align}
-\beta H=&-t\sum_{\langle ij\rangle ,\sigma }\left( c_{i\sigma
}^{\dagger }c_{j\sigma }+c_{j\sigma }^{\dagger }c_{i\sigma
}\right)\label{eq:1} \\
&-U_{0}\sum_{i}n_{i\uparrow }n_{i\downarrow }+\mu
_{0}\sum_{i}n_{i}\:,\nonumber
\end{align}
with $\beta=1/kT$, describing electron conduction on a
$d$-dimensional hypercubic lattice. Here $c_{i\sigma }^{\dagger}$
and $c_{i\sigma} $ respectively are creation and annihilation
operators, obeying anticommutation rules, for an electron with
spin $\sigma =\: \uparrow $ or $\downarrow$ at the site $i$ of the
lattice; $n_{i\sigma }=c_{i\sigma }^{\dagger }c_{i\sigma }$ and
$n_{i}=n_{i\uparrow }+n_{i\downarrow }$ are electron number
operators. Each lattice site can accommodate up to two electrons
with opposite spins. The index $\langle ij\rangle$ denotes
summation over all nearest-neighbor pairs of sites. The three
terms of this Hamiltonian respectively incorporate kinetic energy
(parametrized by the electron hopping strength $t$), on-site
Coulomb repulsion (with coefficient $U_{0}>0$), and chemical
potential $\mu _{0}$. It is convenient for our purposes to
rearrange Eq.(\ref{eq:1}) into an equivalent Hamiltonian by
grouping into a single lattice summation:
\begin{align}
-\beta H=&\sum_{\langle ij\rangle }\left\{ -t\sum_{\sigma }\left(
c_{i\sigma }^{\dagger }c_{j\sigma }+c_{j\sigma }^{\dagger
}c_{i\sigma }\right)\label{eq:2} \right.\\
&\left. -U\left(n_{i\uparrow }n_{i\downarrow
}+n_{j\uparrow}n_{j\downarrow }\right) +\mu \left(
n_{i}+n_{j}\right)\right\}\nonumber\\
\equiv& \sum_{\langle ij\rangle } \left\{-\beta
H(i,j)\right\}\:.\nonumber
\end{align}
The interaction constants are trivially related by
$U=U_{0}/2d,\mu=\mu_{0} /2d$, and we have hereby exhibited the
individual-pair Hamiltonian $-\beta H(i,j)$.

\section{Renormalization-Group Transformation}

\subsection{Exact Formulation in $d=1$}

For $d=1$ (with lattice sites $i=1,2,3,\ldots $), the Hubbard
Hamiltonian in Eq.(\ref{eq:2}) takes the form
\begin{equation}
-\beta H=\sum_{i}\left\{ -\beta H(i,i+1)\right\} ,  \label{eq:3}
\end{equation}
for which an exact renormalization-group transformation can be formulated.  In terms of
matrix elements, this exact transformation is \cite{FalicovBerker}
\begin{multline}
\langle u_{1}u_{3}u_{5}\cdots |e^{-\beta ^{\prime}H^{\prime}}|
v_{1}v_{3}v_{5}\cdots \rangle = \label{eq:4}\\
\sum_{w_{2},w_{4},w_{6},\ldots }\langle
u_{1}w_{2}u_{3}w_{4}u_{5}w_{6}\cdots |e^{-\beta
H}|v_{1}w_{2}v_{3}w_{4}v_{5}w_{6}\cdots \rangle\:,
\end{multline}
where $u_{i}$, $v_{i}$, and $w_{i}$ are state variables for
lattice site $i$. These variables range over the set $\{\circ
,\,\uparrow ,\,\downarrow ,\,\Updownarrow \}$, by which we
represent the no electron, a single electron with spin up, a
single electron with spin down, and doubly occupied states. Here
and below, the quantities referring to the renormalized (rescaled)
system are denoted with a prime. The transformation in
Eq.(\ref{eq:4}) eliminates half of the degrees of freedom in the
system, while exactly preserving the partition function
($Z^{\prime }=Z$). However, the transformation cannot be readily
implemented, due to the non-commutativity of the operators in the
Hamiltonian.

\subsection{Approximation in $d=1$}

The renormalization-group transformation formulated in Sec.IIIA is implemented approximately, as follows:
\begin{equation}
\label{eq:5} \begin{split} \mbox{Tr}_{\mbox{\tiny even}}e^{-\beta
H} =&\mbox{Tr}_{\mbox{\tiny
even}}e^{\sum_{i}\left\{ -\beta H(i,i+1)\right\} }\\
=&\mbox{Tr}_{\mbox{\tiny even}} e^{\sum_{i}^{\mbox{\tiny
even}}\left\{ -\beta H(i-1,i)-\beta H(i,i+1) \right\} }\\
\simeq& \prod_{i}^{\mbox{\tiny even}}\mbox{Tr}_{i}e^{\left\{
-\beta H(i-1,i)-\beta H(i,i+1)\right\} }\\
=&\prod_{i}^{\mbox{\tiny
even}}e^{-\beta ^{\prime }H^{\prime }(i-1,i+1)}\\
\simeq& e^{\sum_{i}^{\mbox{\tiny even}}\left\{ -\beta ^{\prime
}H^{\prime }(i-1,i+1)\right\} } =e^{-\beta ^{\prime }H^{\prime }}.
\end{split}
\end{equation}
In the two approximate steps, marked by $\simeq$ in Eq.(\ref{eq:5}),
we ignore the non-commutation of operators separated beyond three
consecutive sites of the unrenormalized system. Since each of these
two steps involves the same approximation but in opposite
directions, some mutual compensation can be expected. The success of
this approximation at predicting finite-temperature behavior has
been verified in earlier studies of quantum spin systems
\cite{SuzTak,TakSuz}.

The algebraic content of the renormalization-group mapping can be extracted from Eq.(\ref{eq:5}) as
\begin{equation}
e^{-\beta ^{\prime }H^{\prime }(i,k)}=\mbox{Tr}_{j}e^{-\beta H(i,j)-\beta
H(j,k)},
\label{eq:6}
\end{equation}
where $i,j,k$ are three consecutive sites of the unrenormalized system.
The operators $-\beta ^{\prime }H^{\prime }(i,k)$ and $-\beta H(i,j)-\beta
H(j,k)$ act on the space of two-site and three-site states
respectively, so that, in terms of matrix elements,
\begin{multline}
\langle u_{i}v_{k}|e^{-\beta ^{\prime }H^{\prime }(i,k)}|\bar{u}_{i}^{{}}%
\bar{v}_{k}^{{}}\rangle = \label{eq:7}\\
\sum_{w_{j}}\langle u_{i}\,w_{j}\,v_{k}|e^{-\beta H(i,j)-\beta
H(j,k)}|\bar{u}_{i}\,w_{j}\,\bar{v}_{k}^{{}}\rangle \:,
\end{multline}
where $u_{i},w_{j},v_{k},\bar{u}_{i},\bar{v}_{k}^{{}}$ are
single-site state variables. Eq.(\ref{eq:7}) indicates the
contraction of a $64\times 64$ matrix on the right into a
$16\times 16$ matrix on the left. This is greatly simplified by
the use of two- and three-site basis states that block-diagonalize
respectively the left and right sides of Eq.(\ref{eq:7}). These
basis states are the eigenstates of total particle number, total
spin magnitude, total spin $z$-component, and parity. We denote
the set of 16 two-site eigenstates by $\{|\phi _{p}\rangle \}$ and
the set of 64 three-site eigenstates by $\{|\psi _{q}\rangle \}$,
and list them in Tables I and II. \ \ Eq.(\ref{eq:7}) is rewritten
as
\begin{multline}
\langle \phi _{p}|e^{-\beta ^{\prime }H^{\prime }(i,k)}|\phi _{\bar{p}%
}\rangle = \label{eq:8}\\
\sum_{\substack{u,v,\bar{u},\\ \bar{v},w}}
\sum_{\substack{q,\bar{q}}} \langle\phi _p|u_iv_k\rangle \langle
u_iw_jv_k|\psi_q\rangle \langle \psi _q|e^{-\beta H(i,j)-\beta
H(j,k)}|\psi _{\bar{q}}\rangle\cdot \\
\langle \psi_{\bar{q}}|\bar{u}_iw_j\bar{v}_k\rangle \langle
\bar{u}_i\bar{v}_k|\phi _{\bar{p}}\rangle\:.
\end{multline}
In the above equation, with the eigenstates shown in Tables I and
II, the largest block in $\langle \phi _{p}|e^{-\beta ^{\prime
}H^{\prime }(i,k)}|\phi _{\bar{p}}\rangle$ is $2\times 2$ and the
largest block in $\langle \psi _{q}|e^{-\beta H(i,j)-\beta
H(j,k)}|\psi _{\bar{q}}\rangle$ is $4\times 4$. (In previous work
\cite{Hubshort}, some matrix elements in these blocks were
incorrectly derived). Eq.(\ref{eq:8}) yields eleven independent
elements for the matrix $\langle \phi _{p}|e^{-\beta ^{\prime
}H^{\prime }(i,k)}|\phi _{\bar{p}}\rangle$ of the renormalized
system. These we label $\gamma_p$, as shown in Appendix A. The
values of the $\gamma_p$ in terms of the matrix elements of the
unrenormalized system, dictated by the right-hand side of
Eq.(\ref{eq:8}), are also given in Appendix A.

\begin{table}[h]
\begin{tabular}{|c|c|c|c|c|}
 \hline
  $n$ & $p$ & $s$ & $m_s$ & Two-site basis states\\
  \hline
  $0$ & $+$ & $0$ & $0$ &$|\phi_{1}\rangle=|\circ\circ\rangle$ \\
  \hline
  $1$ & $+$ & $1/2$ & $1/2$ &$|\phi_{2}\rangle=\frac{1}{\sqrt{2}}\{|\uparrow
  \circ\rangle+|\circ\uparrow\rangle\}$\\ \hline
  $1$ & $-$ & $1/2$ & $1/2$ &$|\phi_{4}\rangle=\frac{1}{\sqrt{2}}\{|\uparrow
  \circ\rangle-|\circ\uparrow\rangle\}$\\ \hline
  $2$ & $+$ & $0$ & $0$ &
  $|\phi_{6}\rangle=\frac{1}{\sqrt{2}}\{|\Updownarrow\circ\rangle+|\circ\Updownarrow
  \rangle\}$\\ \hline
  $2$ & $-$ & $0$ & $0$ &
  $|\phi_{7}\rangle=\frac{1}{\sqrt{2}}\{|\Updownarrow\circ\rangle-|\circ\Updownarrow
  \rangle\},$\\
  & & & & $|\phi_{8}\rangle=\frac{1}{\sqrt{2}}\{|\uparrow\downarrow\rangle
  -|\downarrow\uparrow\rangle\}$\\ \hline
  $2$ & $+$ & $1$ & $1$ &
  $|\phi_{9}\rangle=|\uparrow\uparrow\rangle$\\ \hline
  $2$ & $+$ & $1$ & $0$ &
  $|\phi_{10}\rangle=\frac{1}{\sqrt{2}}\{|\uparrow\downarrow\rangle+|\downarrow
  \uparrow\rangle\}$\\ \hline
  $3$ & $+$ & $1/2$ & $1/2$ &$|\phi_{12}\rangle=\frac{1}{\sqrt{2}}\{
  |\Updownarrow\uparrow\rangle+|\uparrow\Updownarrow\rangle\}$\\ \hline
  $3$ & $-$ & $1/2$ & $1/2$ &$|\phi_{14}\rangle=\frac{1}{\sqrt{2}}\{
  |\Updownarrow\uparrow\rangle-|\uparrow\Updownarrow\rangle\}$\\ \hline
  $4$ & $+$ & $0$ & $0$ &$|\phi_{16}\rangle=|\Updownarrow\Updownarrow\rangle$ \\
  \hline
\end{tabular}
\caption{The two-site basis states used in the derivation of the
recursion relations, in Eq. (8).  In these basis states,
$e^{-\beta^\prime H^\prime (i,k)}$ is diagonal, with the exception
of a $2\times 2$ block involving $|\phi_{6}\rangle$ and
$|\phi_{8}\rangle$.  The corresponding particle number ($n$),
parity ($p$), total spin ($s$), and total spin $z$-component
($m_s$) quantum numbers are also given.  The states
$|\phi_{3}\rangle$, $|\phi_{5}\rangle$, $|\phi_{11}\rangle$,
$|\phi_{13}\rangle$, $|\phi_{15}\rangle$ are obtained by spin
reversal from $|\phi_{2}\rangle$, $|\phi_{4}\rangle$,
$|\phi_{9}\rangle$, $|\phi_{12}\rangle$, $|\phi_{14}\rangle$,
respectively.}
\end{table}

\begin{table}
\begin{tabular}{|c|c|c|c|c|}
 \hline
  $n$ & $p$ & $s$ & $m_s$ & Three-site basis states\\
  \hline
  $0$ & $+$ & $0$ & $0$ &$|\psi_{1}\rangle=|\circ\circ\,\circ\rangle$ \\
  \hline
  $1$ & $+$ & $1/2$ & $1/2$ &$|\psi_{2}\rangle=|\circ
  \uparrow
  \circ\rangle,\: |\psi_{3}\rangle=\frac{1}{\sqrt{2}}\{|\uparrow
  \circ\,\circ\rangle+|\circ\,\circ\uparrow\rangle\}$\\ \hline
  $1$ & $-$ & $1/2$ & $1/2$ &$|\psi_{6}\rangle=\frac{1}{\sqrt{2}}\{|\uparrow
  \circ\,\circ\rangle-|\circ\,\circ\uparrow\rangle\}$\\ \hline
$2$ & $+$ & $0$ & $0$ &
  $|\psi_{8}\rangle=\frac{1}{2}\{|\uparrow\downarrow\circ\rangle-
  |\downarrow\uparrow\circ\rangle-|\circ\uparrow\downarrow\rangle+
  |\circ\downarrow\uparrow\rangle\},$\\
  &&&&$|\psi_{9}\rangle=|\circ\Updownarrow\circ\rangle, \:
|\psi_{10}\rangle=\frac{1}{\sqrt{2}}\{|\Updownarrow\circ\,\circ\rangle+|\circ\,\circ\Updownarrow
  \rangle\}$\\ \hline
   $2$ & $-$ & $0$ & $0$ &
  $|\psi_{11}\rangle=\frac{1}{2}\{|\uparrow\downarrow\circ\rangle-
  |\downarrow\uparrow\circ\rangle+|\circ\uparrow\downarrow\rangle-
  |\circ\downarrow\uparrow\rangle\},$\\
  &&&&$|\psi_{12}\rangle=\frac{1}{\sqrt{2}}\{|\uparrow\circ\downarrow\rangle-|\downarrow\circ\uparrow
  \rangle\},$\\
  &&&&$|\psi_{13}\rangle=\frac{1}{\sqrt{2}}\{|\Updownarrow\circ\,\circ\rangle-|\circ\,\circ\Updownarrow
  \rangle\}$\\ \hline
  $2$ & $+$ & $1$ & $1$ &
  $|\psi_{14}\rangle=|\uparrow\circ\uparrow\rangle,\:
  |\psi_{15}\rangle=\frac{1}{\sqrt{2}}\{|\uparrow\uparrow\circ\rangle+|\circ\uparrow\uparrow
  \rangle\}$\\ \hline
  $2$ & $+$ & $1$ & $0$ &
  $|\psi_{16}\rangle=\frac{1}{2}\{|\uparrow\downarrow\circ\rangle+
  |\downarrow\uparrow\circ\rangle+|\circ\uparrow\downarrow\rangle+
  |\circ\downarrow\uparrow\rangle\},$\\
  &&&& $|\psi_{17}\rangle=\frac{1}{\sqrt{2}}
  \{|\uparrow\circ\downarrow\rangle+|\downarrow\circ\uparrow
  \rangle\}$\\ \hline
  $2$ & $-$ & $1$ & $1$ &
  $|\psi_{20}\rangle=\frac{1}{\sqrt{2}}\{|\uparrow\uparrow\circ\rangle-|\circ\uparrow\uparrow
  \rangle\}$\\ \hline
  $2$ & $-$ & $1$ & $0$ &
  $|\psi_{21}\rangle=\frac{1}{2}\{|\uparrow\downarrow\circ\rangle+
  |\downarrow\uparrow\circ\rangle-|\circ\uparrow\downarrow\rangle-
  |\circ\downarrow\uparrow\rangle\}$\\ \hline
  $3$ & $+$ & $1/2$ & $1/2$ &
  $|\psi_{23}\rangle=\frac{1}{\sqrt{6}}\{2|\uparrow\downarrow\uparrow\rangle-|\uparrow\uparrow
  \downarrow\rangle-|\downarrow\uparrow\uparrow\rangle\},$\\
  &&&&$|\psi_{24}\rangle=\frac{1}{\sqrt{2}}\{|\uparrow\Updownarrow\circ\rangle+|\circ\Updownarrow
  \uparrow\rangle\},$\\
  &&&&$|\psi_{25}\rangle=\frac{1}{\sqrt{2}}\{|\uparrow\circ\Updownarrow\rangle+|\Updownarrow\circ
  \uparrow\rangle\},$\\
  &&&&$|\psi_{26}\rangle=\frac{1}{\sqrt{2}}\{|\Updownarrow\uparrow\circ\rangle
  +|\circ\uparrow\Updownarrow\rangle\}$\\ \hline
  $3$ & $-$ & $1/2$ & $1/2$ &
  $|\psi_{31}\rangle=\frac{1}{\sqrt{2}}\{|\uparrow\uparrow
  \downarrow\rangle-|\downarrow\uparrow\uparrow\rangle\},$\\
  &&&&$|\psi_{32}\rangle=\frac{1}{\sqrt{2}}\{|\uparrow\Updownarrow\circ\rangle-|\circ\Updownarrow
  \uparrow\rangle\},$\\
  &&&&$|\psi_{33}\rangle=\frac{1}{\sqrt{2}}\{|\uparrow\circ\Updownarrow\rangle-|\Updownarrow\circ
  \uparrow\rangle\},$\\
  &&&&$|\psi_{34}\rangle=\frac{1}{\sqrt{2}}\{|\Updownarrow\uparrow\circ\rangle
  -|\circ\uparrow\Updownarrow\rangle\}$\\ \hline
  $3$ & $+$ & $3/2$ & $3/2$ &
  $|\psi_{39}\rangle=|\uparrow\uparrow\uparrow\rangle$ \\ \hline
  $3$ & $+$ & $3/2$ & $1/2$ &
  $|\psi_{40}\rangle=\frac{1}{\sqrt{3}}\{|\uparrow\downarrow\uparrow\rangle+|\uparrow\uparrow
  \downarrow\rangle+|\downarrow\uparrow\uparrow\rangle\}$ \\ \hline
  $4$ & $+$ & $0$ & $0$ &
  $|\psi_{43}\rangle=|\Updownarrow\circ\Updownarrow\rangle,\:|\psi_{44}\rangle=\frac{1}{\sqrt{2}}\{|\Updownarrow\Updownarrow\circ\rangle
  +|\circ\Updownarrow\Updownarrow\rangle\},$\\
  &&&&$|\psi_{45}\rangle=\frac{1}{2}\{|\uparrow\downarrow\Updownarrow\rangle-
  |\downarrow\uparrow\Updownarrow\rangle-|\Updownarrow\uparrow\downarrow\rangle+
  |\Updownarrow\downarrow\uparrow\rangle\}$\\
  \hline
  $4$ & $-$ & $0$ & $0$ &
  $|\psi_{46}\rangle=\frac{1}{2}\{|\downarrow\uparrow\Updownarrow\rangle
  -|\uparrow\downarrow\Updownarrow\rangle-|\Updownarrow\uparrow\downarrow\rangle+
  |\Updownarrow\downarrow\uparrow\rangle\},$\\
  &&&&$|\psi_{47}\rangle=\frac{1}{\sqrt{2}}\{|\uparrow\Updownarrow\downarrow\rangle-
  |\downarrow\Updownarrow\uparrow\rangle\},$\\
  &&&&$|\psi_{48}\rangle=\frac{1}{\sqrt{2}}\{|\Updownarrow\Updownarrow\circ\rangle
  -|\circ\Updownarrow\Updownarrow\rangle\}$ \\
  \hline
  $4$ & $+$ & $1$ & $1$ &
  $|\psi_{49}\rangle=|\uparrow\Updownarrow\uparrow\rangle,\:
  |\psi_{50}\rangle=\frac{1}{\sqrt{2}}\{|\uparrow\uparrow\Updownarrow\rangle
  +|\Updownarrow\uparrow\uparrow
  \rangle\}$\\ \hline
  $4$ & $+$ & $1$ & $0$ &
  $|\psi_{51}\rangle=\frac{1}{2}\{|\uparrow\downarrow\Updownarrow\rangle+
  |\downarrow\uparrow\Updownarrow\rangle+|\Updownarrow\uparrow\downarrow\rangle+
  |\Updownarrow\downarrow\uparrow\rangle\},$\\
  &&&&$|\psi_{52}\rangle=\frac{1}{\sqrt{2}}
  \{|\uparrow\Updownarrow\downarrow\rangle+|\downarrow\Updownarrow\uparrow
  \rangle\}$\\ \hline
  $4$ & $-$ & $1$ & $1$ &
  $|\psi_{55}\rangle=\frac{1}{\sqrt{2}}\{|\uparrow\uparrow\Updownarrow\rangle-
  |\Updownarrow\uparrow\uparrow
  \rangle\}$\\ \hline
  $4$ & $-$ & $1$ & $0$ &
  $|\psi_{56}\rangle=\frac{1}{2}\{|\uparrow\downarrow\Updownarrow\rangle+
  |\downarrow\uparrow\Updownarrow\rangle-|\Updownarrow\uparrow\downarrow\rangle-
  |\Updownarrow\downarrow\uparrow\rangle\}$\\ \hline
   $5$ & $+$ & $1/2$ & $1/2$ &$|\psi_{58}\rangle=|\Updownarrow\uparrow\Updownarrow\rangle,
  \: |\psi_{59}\rangle=\frac{1}{\sqrt{2}}\{|\uparrow
  \Updownarrow\Updownarrow\rangle+|\Updownarrow\Updownarrow\uparrow\rangle\}$\\ \hline
  $5$ & $-$ & $1/2$ & $1/2$ &$|\psi_{62}\rangle=\frac{1}{\sqrt{2}}\{
  |\Updownarrow\Updownarrow\uparrow\rangle-|\uparrow
  \Updownarrow\Updownarrow\rangle\}$\\ \hline
  $6$ & $+$ & $0$ & $0$ &
  $|\psi_{64}\rangle=|\Updownarrow\Updownarrow\Updownarrow\rangle$\\ \hline
\end{tabular}
\caption{The three-site basis states used in the derivation of the
recursion relations, in Eq. (8).  In these basis states, $e^{-\beta
H (i,j)-\beta H(j,k)}$ is block-diagonal, with the largest blocks
being $4\times 4$ (see Table IV).  The corresponding particle number
($n$), parity ($p$), total spin ($s$), and total spin $z$-component
($m_s$) quantum numbers are also given. The states
$|\phi_{4-5}\rangle$, $|\phi_{7}\rangle$, $|\phi_{18-19}\rangle$,
$|\phi_{22}\rangle$, $|\phi_{27-30}\rangle$, $|\phi_{35-38}\rangle$,
$|\phi_{41-42}\rangle$, $|\phi_{53-54}\rangle$, $|\phi_{57}\rangle$,
$|\phi_{60-61}\rangle$, $|\phi_{63}\rangle$ are obtained by spin
reversal from $|\phi_{2-3}\rangle$, $|\phi_{6}\rangle$,
$|\phi_{14-15}\rangle$, $|\phi_{20}\rangle$, $|\phi_{23-26}\rangle$,
$|\phi_{31-34}\rangle$, $|\phi_{39-40}\rangle$,
$|\phi_{49-50}\rangle$, $|\phi_{55}\rangle$, $|\phi_{58-59}\rangle$,
$|\phi_{62}\rangle$, respectively.}
\end{table}

\subsection{Hamiltonian Closed Form under\\ the Renormalization-Group Transformation}

Since eleven interaction strengths can be independently fixed by the
eleven $\gamma_p$, the Hamiltonian $-\beta^\prime H^\prime$ which is
embodied in Appendix A has a more general form than that of the
Hubbard Hamiltonian in Eq.(\ref{eq:2}). This generalized form of the
pair Hamiltonian is
\begin{equation}
\label{eq:9}
\begin{split}
-&\beta H(i,j) = \\
&-\sum_{\substack{\sigma}}\:\left[\,t_0
\,h_{i\,-\sigma}h_{j\,-\sigma}
+t_1(h_{i\,-\sigma}n_{j\,-\sigma}+n_{i\,-\sigma}h_{j\,-\sigma})\right.\\
&\left. +t_2\,n_{i\,-\sigma}n_{j\,-\sigma}\right]
\left(c^\dagger_{i\sigma}c_{j\sigma}+c^\dagger_{j\sigma}c_{i\sigma}\right)\\
&-t_x \left(c^\dagger_{i\uparrow}
c_{j\uparrow}c^\dagger_{i\downarrow}
c_{j\downarrow}+c^\dagger_{j\uparrow}
c_{i\uparrow}c^\dagger_{j\downarrow}
c_{i\downarrow}\right)\\
&-U\left(n_{i\uparrow}n_{i\downarrow}+n_{j\uparrow}n_{j\downarrow}\right)+
\mu\left(n_i+n_j\right) +J\,\vec{S}_i\cdot\vec{S}_j\\
& +V_2 n_i n_j + V_3\left(n_{i\uparrow}n_{i\downarrow}n_j +n_i
n_{j\uparrow}n_{j\downarrow}\right)\\
&+V_4
n_{i\uparrow}n_{i\downarrow}n_{j\uparrow}n_{j\downarrow}\:+\:G\:,
\end{split}
\end{equation}
where $h_{i\sigma} \equiv 1-n_{i\sigma}$ is the hole (vacancy)
operator and $\vec{S}_i = \sum_{\sigma,\bar{\sigma}}
c^\dagger_{i\sigma} \vec{s}_{\sigma\bar{\sigma}}
c_{i\bar{\sigma}}$, with $\vec{s}_{\sigma\bar{\sigma}}$ the vector
of Pauli spin matrices, is the spin operator at site i.  In
general, the Hubbard Hamiltonian, after one renormalization-group
transformation, maps onto this generalized Hamiltonian, which has
a form that stays closed under further renormalization-group
transformations.

The kinetic energy part of the Hamiltonian in Eq.(\ref{eq:9}) distinguishes the
four types of nearest-neighbor hopping events:
{\bf i) vacancy hopping} (the $t_0$ term): a vacancy (hole) hopping against a
background of single-electron occupancy (half-filling);
{\bf ii) pair breaking or pair making} (the $t_1$ term): doubly occupied and completely
unoccupied nearest-neighbor sites reverting to half-filling, or the reverse process;
{\bf iii) pair hopping} (the $t_2$ term): a pair hopping against a background of
half-filling;
{\bf iv) vacancy - pair interchange} (the $t_x$ term): doubly occupied and completely
unoccupied nearest-neighbor sites exchanging positions.

The generalized Hamiltonian of Eq.(\ref{eq:9}) reduces to the
Hubbard Hamiltonian of Eq.(\ref{eq:2}) for $t_0=t_1=t_2=t$ and
$t_x=J=V_2=V_3=V_4=G=0\:.$ The renormalization-group flows occur in
the 10-dimensional interaction space of the generalized Hamiltonian;
the 3-dimensional interaction space of the Hubbard Hamiltonian
contains the initial conditions of the renormalization-group flows.

The matrix elements of the renormalized pair Hamiltonian
$-\beta^\prime H^\prime(i,k)$ are given in Table III in terms of the
renormalized interaction constants. Table III allows us to solve for
the renormalized interaction constants in terms of the $\gamma_p$
given in Appendix A:
\begin{gather}
t_0^\prime = \frac{1}{2}\ln\frac{\gamma_4}{\gamma_2},\qquad
t_1^\prime = u \frac{\gamma_{0}}{\gamma_{8}-\gamma_{6}},\nonumber\\[5pt]
t_2^\prime = \frac{1}{2}\ln\frac{\gamma_{12}}{\gamma_{14}},\qquad
t_x^\prime = \frac{1}{2}\left(u -
v+\ln\gamma_7\right),\nonumber\\[5pt]
U^\prime = \frac{1}{2}\left(u - v +
\ln\frac{\gamma_2^2\gamma_4^2}{\gamma_1^2\gamma_7}\right),\qquad
\mu^\prime = \frac{1}{2}\ln\frac{\gamma_2\gamma_4}{\gamma_1^2},\nonumber\\[5pt]
J^\prime = -u-v+\ln\gamma_9,\qquad V_2^\prime =
\frac{1}{4}\ln\frac{\gamma_1^4\gamma_9^3}{\gamma_2^4\gamma_4^4}
+\frac{1}{4}(u+v),\nonumber\\[5pt]
V_3^\prime =
\frac{1}{2}\ln\frac{\gamma_2^3\gamma_4^3\gamma_{12}\gamma_{14}}
{\gamma_1^2\gamma_7\gamma_9^3} -v,\qquad V_4^\prime =
\ln\frac{\gamma_1\gamma_7\gamma_9^3\gamma_{16}}
{\gamma_2^2\gamma_4^2\gamma_{12}^2\gamma_{14}^2}+2v,\nonumber\\[5pt]
\label{eq:10}G^\prime=\ln\gamma_1,
\end{gather}
where
\begin{gather*}
v = \frac{1}{2}\ln\left(\gamma_6\gamma_{8}-\gamma_0^2\right),\\[5pt]
u = \frac{\gamma_{8}-\gamma_6}
{\sqrt{\left(\gamma_{8}-\gamma_6\right)^2+4\gamma_0^2}} \cosh^{-1}
\left(\frac{\gamma_{8}+\gamma_6}{2e^v}\right).
\end{gather*}

\begingroup
\squeezetable
\begin{table}
\begin{gather*}
\begin{array}{|c||c|c|c|c|c|c|} \hline -\beta^\prime
H^\prime(i,k) & \phi_{1} &\phi_{2} &\phi_{4} & \phi_{7} &
\phi_{9} &\phi_{10}\\
\hhline{|=#=|=|=|=|=|=|} \phi_{1} & G^\prime & \multicolumn{5}{c|}{}\\
\cline{1-3} \phi_{2} && \parbox{0.35in}{\centering $-t_0^\prime+\mu^\prime+G^\prime$} &\multicolumn{4}{c|}{0}\\
\cline{1-1}\cline{3-4} \phi_{4} & \multicolumn{2}{c|}{}
&\parbox{0.35in}{\centering $t_0^\prime+\mu^\prime+G^\prime$} &\multicolumn{3}{c|}{}\\
\cline{1-1}\cline{4-5} \phi_{7} & \multicolumn{3}{c|}{} &
\multicolumn{1}{c|}{\parbox{0.4in}{\centering
$t_x^\prime-U^\prime+2\mu^\prime+G^\prime$}} &
\multicolumn{2}{c|}{}\\
\cline{1-1}\cline{5-6} \phi_{9} & \multicolumn{4}{c|}{0}
&\multicolumn{1}{c|}{\parbox{0.4in}{\centering
$2\mu^\prime+\frac{1}{4}J^\prime +V_2^\prime+G^\prime$}}&\\
\cline{1-1}\cline{6-7} \phi_{10} & \multicolumn{5}{c|}{}
& \parbox{0.4in}{\centering $2\mu^\prime+\frac{1}{4}J^\prime+V_2^\prime+G^\prime$}\\
\hline
\end{array}\\
\begin{array}{|c||c|c|} \hline -\beta^\prime
H^\prime(i,k) & \phi_{6} & \phi_{8} \\
\hhline{|=#=|=|} \phi_{6} &
-t_x^\prime-U^\prime+2\mu^\prime+G^\prime
&2t_1^\prime\\
\cline{1-3} \phi_{8} &2t_1^\prime
&2\mu^\prime-\frac{3}{4}J^\prime+V_2^\prime+G^\prime
\\
\hline
\end{array}\\
\begin{array}{|c||c|c|c|} \hline -\beta^\prime H^\prime(i,k)
& \phi_{12} &\phi_{14} & \phi_{16}\\
\hhline{|=#=|=|=|} \phi_{12} &
\parbox{0.75in}{\centering
$t_2^\prime-U^\prime+3\mu^\prime$\\
$+2V_2^\prime+V_3^\prime+G^\prime$}
&\multicolumn{2}{r|}{}\\
\cline{1-3} \phi_{14} & \multicolumn{1}{c|}{}
&\parbox{0.75in}{\centering $-t_2^\prime-U^\prime+3\mu^\prime$\\ $+2V_2^\prime+V_3^\prime+G^\prime$} & 0\\
\cline{1-1}\cline{3-4} \phi_{16} & \multicolumn{2}{c|}{0}
&\parbox{0.8in}{\centering $-2U^\prime+4\mu^\prime+4V_2^\prime$\\
$+4V_3^\prime+V_4^\prime
+G^\prime$}\\
\hline
\end{array}
\end{gather*}
\caption{Block-diagonal matrix of the renormalized two-site
Hamiltonian $-\beta^\prime H^\prime(i,k)$.  The Hamiltonian being
invariant under spin-reversal, the spin-flipped matrix elements
are not shown.}
\end{table}
\endgroup

\begingroup
\squeezetable
\begin{table}
\begin{gather*}
\begin{array}{|c||c|}\hline
 & \psi_{1}\\
\hhline{|=#=|} \psi_{1} & 0\\ \hline
\end{array}
\quad
\begin{array}{|c||c|c|}\hline
 & \psi_{2} & \psi_{3}\\
\hhline{|=#=|=|} \psi_{2} & 2\mu & -\sqrt{2} t_0\\
\hline \psi_{3} & -\sqrt{2} t_0 & \mu\\
\hline
\end{array}
\quad
\begin{array}{|c||c|}\hline
 & \psi_{6}\\
\hhline{|=#=|} \psi_{6} & \mu\\ \hline
\end{array}\\
\begin{array}{|c||c|c|c|c|}\hline
 & \psi_{9} & \psi_{10} & \psi_{11} &\psi_{12}\\
\hhline{|=#=|=|=|=|} \psi_{9} & -2U+4\mu & -\sqrt{2}t_x & 2 t_1 & 0\\
\hline \psi_{10} & -\sqrt{2} t_x & -U+2\mu & \sqrt{2} t_1 & 0\\
\hline \psi_{11} & 2 t_1 & \sqrt{2} t_1 & 3\mu-\frac{3}{4}J+V_2 & -\sqrt{2} t_0\\
\hline \psi_{12} & 0 & 0 & -\sqrt{2} t_0 & 2\mu\\ \hline
\end{array}\\
\begin{array}{|c||c|c|}\hline
 & \psi_{8} & \psi_{13}\\
\hhline{|=#=|=|} \psi_{8} & 3\mu-\frac{3}{4} J+V_2 & \sqrt{2}t_1\\
\hline \psi_{13} & \sqrt{2}t_1 & -U+2\mu\\ \hline
\end{array}
\quad
\begin{array}{|c||c|c|}\hline
 & \psi_{14} & \psi_{15}\\
\hhline{|=#=|=|} \psi_{14} & 2\mu & -\sqrt{2} t_0\\
\hline \psi_{15} & -\sqrt{2} t_0 & 3\mu+\frac{1}{4}J+V_2\\
\hline
\end{array}\\
\begin{array}{|c||c|c|}\hline
 & \psi_{16} & \psi_{17}\\
\hhline{|=#=|=|} \psi_{16} & 3\mu+\frac{1}{4} J+V_2 & -\sqrt{2} t_0\\
\hline \psi_{17} & -\sqrt{2} t_0 & 2\mu\\
\hline
\end{array}
\quad
\begin{array}{|c||c|}\hline
 & \psi_{20}\\
\hhline{|=#=|} \psi_{20} & \parbox{0.4in}{\centering $3\mu+\frac{1}{4}J+V_2$}\\
\hline
\end{array}
\quad
\begin{array}{|c||c|}\hline
 & \psi_{21}\\
\hhline{|=#=|} \psi_{21} & \parbox{0.4in}{\centering $3\mu+\frac{1}{4}J+V_2$}\\
\hline
\end{array}\\
\begin{array}{|c||c|c|c|c|c|} \hline
 & \psi_{24} & \psi_{25} &\psi_{26}&\psi_{31}\\
\hhline{|=#=|=|=|=|} \psi_{24} &
\parbox{0.6in}{\centering $-2U+5\mu+2V_2+V_3$} &
-t_x & t_2 & t_1\\
\hline \psi_{25}  & -t_x & -U+3\mu & -t_0
& t_1\\
\hline \psi_{26}  & t_2 & -t_0 & \parbox{0.5in}{\centering $-U+4\mu+2V_2+V_3$} & 0\\
\hline \psi_{31}  & t_1 &
t_1 & 0 & \parbox{0.5in}{\centering $4\mu+2V_2$}\\
\hline
\end{array}\\
\begin{array}{|c||c|c|c|c|} \hline
 & \psi_{23} & \psi_{32} & \psi_{33} &\psi_{34}\\
\hhline{|=#=|=|=|=|} \psi_{23} & 4\mu-J+2V_2 & -\sqrt{3} t_1 & -\sqrt{3} t_1 & 0\\
\hline \psi_{32} & -\sqrt{3} t_1 & \parbox{0.6in}{\centering $-2U+5\mu+2V_2+V_3$} & -t_x & t_2\\
\hline \psi_{33} & -\sqrt{3} t_1 & -t_x & -U+3\mu & t_0\\
\hline \psi_{34} & 0 & t_2 & t_0 & \parbox{0.6in}{\centering $-U+4\mu+2V_2+V_3$}\\
\hline
\end{array}\\
\begin{array}{|c||c|}\hline
 & \psi_{39}\\
\hhline{|=#=|} \psi_{39} & 4\mu+\frac{1}{2}J+2V_2\\ \hline
\end{array} \quad
\begin{array}{|c||c|} \hline
 & \psi_{40}\\
\hhline{|=#=|} \psi_{40} & 4\mu+\frac{1}{2}J+2V_2\\
\hline
\end{array}\\
\begin{array}{|c||c|c|c|c|} \hline
 & \psi_{43} & \psi_{44} & \psi_{46} &\psi_{47}\\
\hhline{|=#=|=|=|=|} \psi_{43} & -2U+4\mu & -\sqrt{2}t_x & -2t_1 & 0\\
\hline \psi_{44} & -\sqrt{2}t_x & \parbox{0.7in}{\centering $-3U+6\mu+4V_2+4V_3+V_4$} & -\sqrt{2}t_1 & 0\\
\hline \psi_{46} & -2t_1 & -\sqrt{2}t_1 & \parbox{0.7in}{\centering $-U+5\mu-\frac{3}{4}J+3V_2+V_3$} & -\sqrt{2}t_2\\
\hline \psi_{47} & 0 & 0 & -\sqrt{2}t_2 & \parbox{0.6in}{\centering $-2U+6\mu+4V_2+2V_3$}\\
\hline
\end{array}\\
\begin{array}{|c||c|c|} \hline
 & \psi_{45} & \psi_{48}\\
\hhline{|=#=|=|} \psi_{45} & -U+5\mu-\frac{3}{4}J+3V_2+V_3 & -\sqrt{2} t_1\\
\hline \psi_{48} & -\sqrt{2} t_1 & -3U+6\mu+4V_2+4V_3+V_4\\
\hline
\end{array}\\
\begin{array}{|c||c|c|}\hline
 & \psi_{49} & \psi_{50}\\
\hhline{|=#=|=|} \psi_{49} & -2U+6\mu+4V_2+2V_3 & \sqrt{2} t_2\\
\hline \psi_{50} & \sqrt{2} t_2 & -U+5\mu+\frac{1}{4}J+3V_2+V_3\\
\hline
\end{array}\\
\begin{array}{|c||c|c|} \hline
 & \psi_{51} & \psi_{52}\\
\hhline{|=#=|=|} \psi_{51} & -U+5\mu+\frac{1}{4} J+3V_2+V_3 & \sqrt{2} t_2\\
\hline \psi_{52} & \sqrt{2} t_2 & -2U+6\mu+4V_2+2V_3\\
\hline
\end{array}\\
\begin{array}{|c||c|} \hline
 & \psi_{55}\\
\hhline{|=#=|} \psi_{55} & \parbox{0.8in}{\centering $-U+5\mu+\frac{1}{4}J+3V_2+V_3$}\\
\hline
\end{array}
\quad
\begin{array}{|c||c|} \hline
 & \psi_{56}\\
\hhline{|=#=|} \psi_{56} & \parbox{0.8in}{\centering $-U+5\mu+\frac{1}{4}J+3V_2+V_3$}\\
\hline
\end{array}\\
\begin{array}{|c||c|c|} \hline
 & \psi_{58} & \psi_{59}\\
\hhline{|=#=|=|} \psi_{58} & -2U+6\mu+4V_2+2V_3 & \sqrt{2} t_2\\
\hline \psi_{59} & \sqrt{2} t_2 & -3U+7\mu+6V_2+5V_3+V_4\\ \hline
\end{array}\\
\begin{array}{|c||c|} \hline
 & \psi_{62}\\
\hhline{|=#=|} \psi_{62} & \parbox{1in}{\centering $-3U+7\mu+6V_2+5V_3+V_4$}\\
\hline
\end{array}
\quad
\begin{array}{|c||c|} \hline
 & \psi_{64}\\
\hhline{|=#=|} \psi_{64} & \parbox{1in}{\centering
$-4U+8\mu+8V_2+8V_3+2V_4$}\\ \hline
\end{array}
\end{gather*}
\caption{Diagonal matrix blocks of the unrenormalized three-site
Hamiltonian $-\beta H(i,j)-\beta H(j,k)$.  The Hamiltonian being
invariant under spin-reversal, the spin-flipped matrix elements
are not shown.  The additive constant contribution $2G$, occurring
at the diagonal terms, is also not shown.}
\end{table}
\endgroup

This completes the determination of our renormalization-group
transformation, whose flows in the ten-dimensional interaction
space ($t_0,$\,$t_1,$\,$t_2,$\,$t_x,$
\,$U,$\,$\mu,$\,$J,$\,$V_2,$\,$V_3,$\,$V_4$) are to be analyzed.
($G$ is an additive constant not influencing the flows of the 10
other interaction constants.  However, for expectation value
calculations, its derivatives must be included in
Eq.(\ref{eq:14}).)

\subsection{$d=1$ Renormalization-Group Transformation}

The transformation described above is the removal (decimation) of every other site
in a linear array.  This decimation produces the mapping of a Hamiltonian with interaction
constants $\mathbf{K}=(t_0,t_1,t_2,t_x,U,\mu,J,V_2,V_3,V_4,G)$ onto another Hamiltonian
with interaction constants

\begin{equation}\label{eq:12}
\mathbf{K^\prime} = \mathbf{R}(\mathbf{K}).
\end{equation}
The function $\mathbf{R}$ is calculated as follows:

(1) The matrix elements of $-\beta H(i,j)-\beta H(j,k)$ are
determined in the three-site basis $\{\psi_q\}$ given in Table II.
In this basis, this matrix is block-diagonal as shown in Table IV,
with the largest block being $4\times 4$.

(2) The above block-diagonal matrix is exponentiated, yielding the
matrix elements $\langle \psi_q | e^{-\beta H(i,j)-\beta
H(j,k)}|\psi_{\bar{q}}\rangle$ which enter on the right-hand side of
Eq.(\ref{eq:8}). This in turn yields the eleven $\gamma_p$ (as given
in Appendix A).

(3) Using Eqs.(\ref{eq:10}), the interaction constants of the
renormalized Hamiltonian $-\beta^\prime H^\prime(i,k)$, namely
($t^\prime_0$,\,$t^\prime_1$,\,$t^\prime_2$,
\,$t^\prime_x$,\,$U^\prime$,\,$\mu^\prime$,\,$J^\prime$,\,$V^\prime_2$,\,$V^\prime_3$,
\,$V^\prime_4$,\,$G^\prime$) are found.

The initial conditions, for the iterated renormalization-group transformations
that constitute the renormalization-group flow, are the interaction constants of the
Hubbard Hamiltonian, $\mathbf{K}_0 =(t_0=t,\,t_1=t,\, t_2=t,\, t_x=0,\,
U,\,\mu,\,J=0,\,V_2=0,\,V_3=0,\,V_4=0,\,G=0)$.

\subsection{$d>1$ Renormalization-Group Transformation}

The Migdal-Kadanoff approximation procedure \cite{Migdal,Kadanoff}
(which has been remarkably effective in problems as diverse as
lower-critical dimensions for different types of phase transitions;
first- and second-order phase transitions in q-state Potts models;
algebraic order in the $d=2$ XY model; random-field, random-bond,
spin-glass systems; quenched-disorder-induced criticality; etc.) is
used to construct the renormalization-group transformation for
$d>1$.  In the $d$-dimensional hypercubic lattice, a subset of the
nearest-neighbor interactions are ignored, so that a hypercubic
lattice (still $d$-dimensional) is left behind, in which each
lattice point is connected by two consecutive nearest-neighbor
segments of the original lattice. The decimation described above can
then be applied to the site connecting these two segments of the
original lattice, yielding the renormalized nearest-neighbor
couplings between the lattice points of the new hypercubic lattice.
To compensate for the nearest-neighbor interactions that are
ignored, the couplings are multiplied by a factor of $b^{d-1}$ after
decimation, $b=2$ being the length rescaling factor. Thus, the
renormalization-group transformation of Eq.(\ref{eq:12}) in the
previous section generalizes, for $d>1$, to
\begin{equation} \label{eq:13}
\mathbf{K^\prime}=b^{d-1} \mathbf{R}(\mathbf{K}).
\end{equation}

\subsection{Supporting Results}

New global phase diagrams obtained by approximate
renormalization-group transformations are supported by the correct
rendition of all of the special cases of the system solved.  The
Hamiltonian in Eq.(\ref{eq:9}), which is the system presently solved
by approximate recursion relations, reduces in various limits to the
Ising, quantum XY, and quantum Heisenberg spin systems.  Our
recursion relations correctly yield the lower critical-dimensions
$d_l$ of the Ising ($d_l=1$), quantum XY ($d_l=2$), and quantum
Heisenberg ($d_l=2$) spin systems.  For the quantum XY spin system
in $d=2$, this approximation yields the algebraically ordered
Kosterlitz-Thouless low-temperature phase.\cite{SuzTak,TakSuz}  For
the quantum Heisenberg spin system in $d=3$, our recursion relations
yield low-temperature antiferromagnetically (for $J<0$) and
ferromagnetically (for $J>0$) ordered phases, each separated by a
second-order transition from the high-temperature disordered phase.
The antiferromagnetic transition temperature is thus found to be
1.22 times \cite{FalicovBerker} the ferromagnetic transition
temperature, a purely quantum mechanical effect, and to be compared
with the value of 1.13 from series expansion
\cite{RushWood,OitmaaZheng}. Furthermore, as purely off-diagonal
quantum effects, the hopping-induced antiferromagnetism of the $d=3$
Hubbard model is recovered and the scaling of the antiferromagnetic
transition temperature is obtained with an excellent quantitative
agreement, as discussed in Sec.V at Eq.(\ref{eq:15b}) and shown in
Fig.3. In fact, the scaling of the antiferromagnetic transition at
strong-coupling (Fig.~3), as well as the results quoted above, and
the disappearance of the transition at zero coupling (Fig.~4),
indicate the validity of our approximation across the entire
strong-to-weak coupling range. Finally, the Blume-Emery-Griffiths
model is contained in the Hamiltonian of Eq.(\ref{eq:9}) and its
global phase diagram \cite{BerkerWortis} is obtained from our
recursion relations.  All of these results strongly support the
validity of the global calculation here.

\section{Renormalization-Group Analysis:
Global Phase Diagram and\hspace{0.2in} Operator Expectation Values}

From the recursion equations determined in the preceding section,
flows are generated for initial values of $t,\,U$, and $\mu$ in
the Hubbard Hamiltonian.  The renormalization-group
transformation, which constitutes each step of the flow, is
effected numerically. Particular attention has to be given to the
multiplication of small amplitudes with large exponentials, which
can occur in the right-hand side of Eq.(\ref{eq:8}) when
interaction constants become large, causing the computational
difficulties encountered in previous work \cite{Hubshort}.

Each completely stable fixed point, namely sink of the renormalization-group flows,
corresponds to a thermodynamic phase, and the global phase diagram is found by identifying
the basin of attraction for every sink.\cite{BerkerWortis}  The expectation values for the
operators occurring in the Hamiltonian are obtained from the conjugate recursion relations,
\cite{BOP}

\begin{equation}\label{eq:14}
n_\beta = b^{-d} n_\alpha^\prime T_{\alpha\beta}\,,
\end{equation}
with summation over the repeated index $\alpha$ implicit.  The
recursion matrix is
\begin{equation}\label{eq:14ex}
T_{\alpha\beta}=\frac{\partial K^\prime_\alpha}{\partial K_\beta}\,,
\end{equation}
where $K_\alpha$ is an interaction strength, namely a component
in the interaction strength vector $\mathbf{K}$ defined before
Eq.(\ref{eq:12}); $n_\alpha$ is the expectation value of the
operator that occurs in the Hamiltonian with coefficient $K_\alpha$.
Eq.(\ref{eq:14}) is iterated along a trajectory until a phase sink
limit.  The left eigenvector of $T_{\alpha\beta}$ with eigenvalue
$b^d$ gives the expectation values at the phase sink, thereby
completing the calculation of the expectation values of the initial
point of the trajectory.

The observed phase sinks in the calculations for the $d=3$ Hubbard
model --- the details of which are shown in Table V --- have a
property in common: at the sink limit, $t_1$ renormalizes toward
zero.  In the limit $t_1 \to 0$, analytic expressions are derived
to first order in $t_1$ for the matrix elements $\langle\psi_q |
e^{-\beta H(i,j)-\beta H(j,k)}|\psi_{\bar{q}}\rangle$ on the
right-hand side of Eq.(\ref{eq:8}).  This yields, in the
neighborhood of each phase sink, analytic renormalization-group
equations.  The analytic equations provide a useful check on the
accuracy of the numerical calculations, and lead to closed-form
expressions for limiting values of interaction strengths or ratios
of limiting values of interaction strengths.

Flows that start at the boundaries between phases have their own
fixed points, distinguished from phase sinks by having at least
one unstable direction.  After narrowing down onto the boundary
and from there following a flow to the neighborhood of the
unstable fixed point, a Newton-Raphson procedure is used to
exactly locate this unstable fixed point.  Analysis at these fixed
points determines the phase transition properties.  The
expectation values calculated, as described above, at the phase
boundaries allow us to redraw the phase diagram using expectation
values $n_\alpha$ on the axes as well as $t,\, U$, and $\mu$.

The Hamiltonian of Eq.(\ref{eq:9}) is covariant under
particle-hole symmetry ($c^\dagger_{i\sigma} \to c_{i\sigma}$),
which in Hamiltonian space takes the form of a mapping
$\mathbf{\bar{K}}=\mathbf{S}(\mathbf{K})$. The function
$\mathbf{S}$ is given by
\begin{gather}
\bar{t_0}=-t_2,\ \bar{t_1}=-t_1,\ \bar{t_2}=-t_0,\ \bar{t_x}=t_x,
\
\bar{J}=J,\nonumber\\
\bar{U}=U-2V_3-V_4,\ \bar{\mu}= -\mu+U-2V_2-3V_3-V_4,\nonumber\\
\bar{V_2}=V_2+2V_3+V_4,\ \bar{V_3}=-V_3-V_4,\ \bar{V_4}=V_4.
\label{eq:15}
\end{gather}
The subspace that is invariant under $\mathbf{S}$ corresponds to
systems that are invariant under particle-hole exchange, and
therefore are at half-filling: $\langle n_{i} \rangle = 1 = \langle
h_{i} \rangle$.  From Eq.(\ref{eq:15}), this subspace occurs at
$t_0=-t_2,\; t_1=0,\; 2\mu=U-2V_2-V_3,\; 2V_3=-V_4$. For the
original Hubbard Hamiltonian, all points with $\mu_0/U_0=1/2$ are
mapped onto this subspace after the first renormalization-group
step.

The Hubbard phase diagrams are plotted in the next section, for
fixed $U_0/t$, in terms of $1/t$ (a temperature variable) versus
$\mu_0/U_0$ or $\langle n_{i} \rangle$.  Since our
renormalization-group transformation is also covariant under
particle-hole symmetry, the phase diagrams are duly symmetric about
$\mu_0/U_0=1/2$ or $\langle n_{i} \rangle = 1$.

\section{Global Phase Diagram for $d=3$}

\begin{figure*}
\centering

\includegraphics*[scale=0.9]{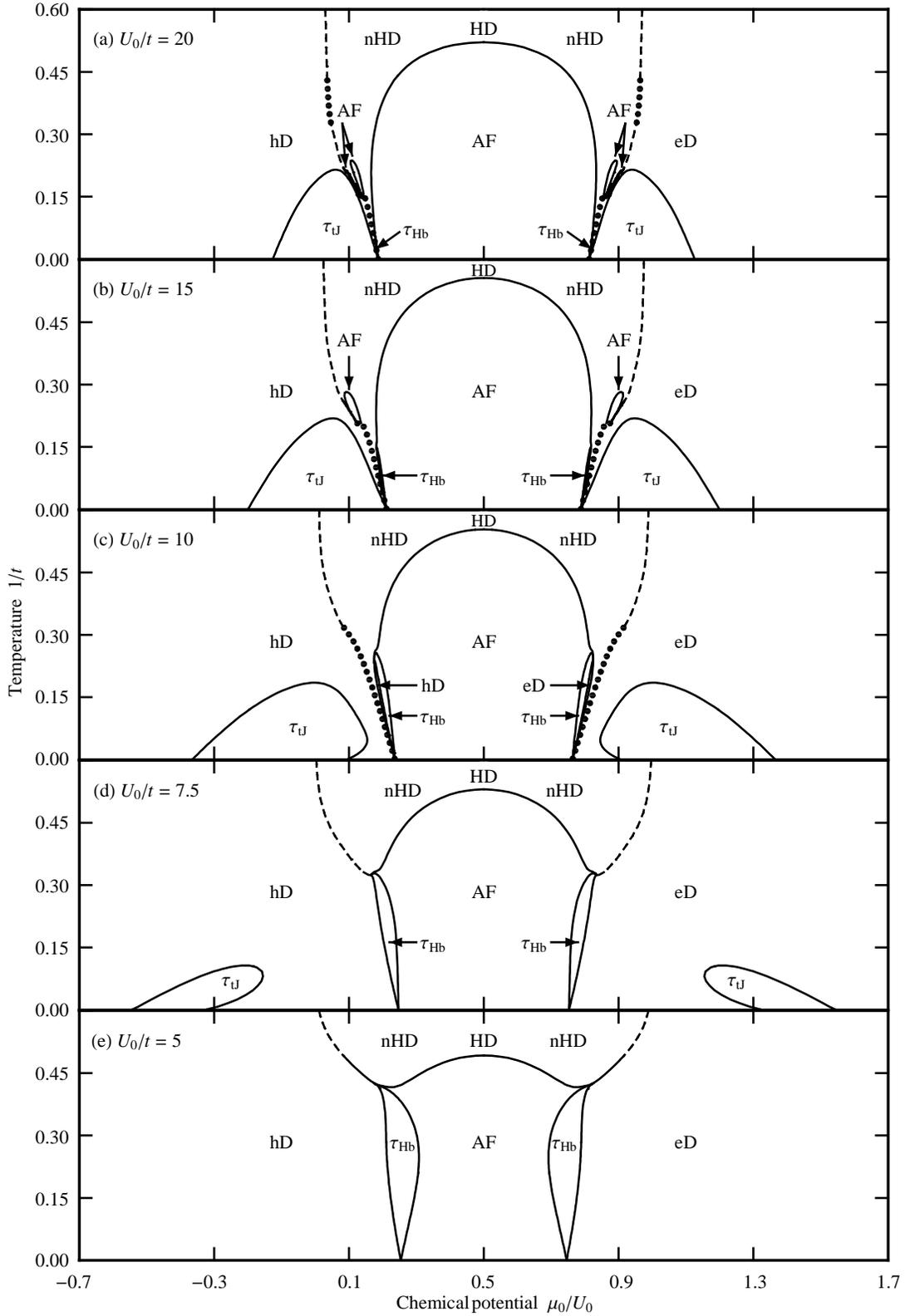}

\caption{$d=3$ Hubbard model phase diagrams in temperature versus
chemical potential. The hole-rich disordered (hD), near-half-filled
disordered (nHD), half-filled disordered (HD), electron-rich
disordered (eD), antiferromagnetic (AF), $\tau_{\text{Hb}}$, and
$\tau_{\text{tJ}}$ phases are seen. The full curves are second-order
phase boundaries, while the dotted curves are first-order
boundaries.  The dashed curves are not phase transitions, but
disorder lines between the near-half-filled disordered and the
hole-rich or electron-rich disordered phases.  The progression (a)
$U_0/t=20$ through (e) $U_0/t = 5$ shows the changing phase diagram
topology from strong to intermediate coupling.  The
$\tau_{\text{tJ}}$ phase, which is prominent at strong coupling,
disappears entirely for $U_0/t \lesssim 6$, and the
$\tau_{\text{Hb}}$ phase is prominent for intermediate couplings. }

\end{figure*}

\begin{figure*}
\centering

\includegraphics*[scale=0.9]{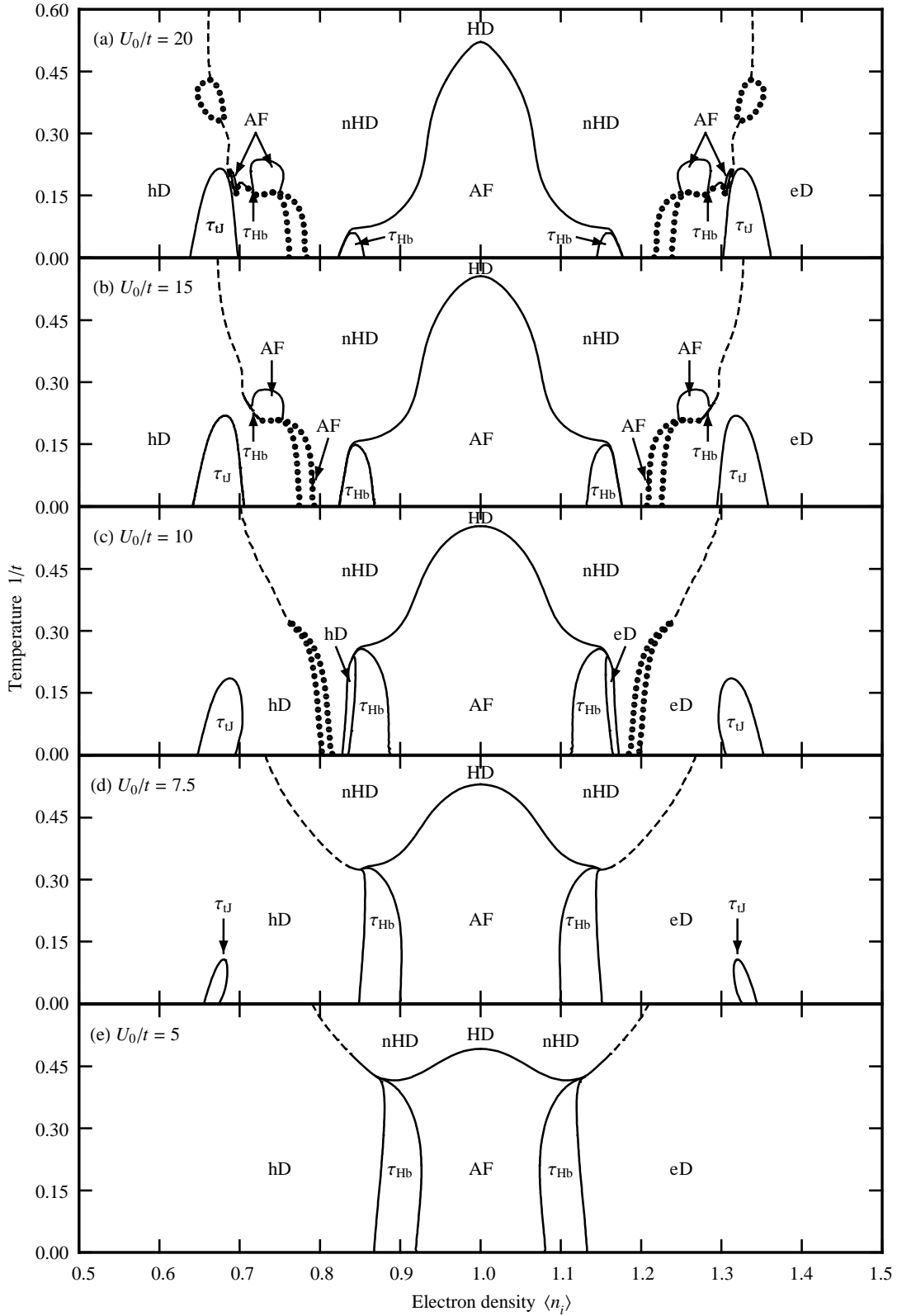}

\caption{$d=3$ Hubbard model phase diagrams in temperature versus
electron density.  The full curves are second-order phase
boundaries.  The coexistence boundaries of first-order transitions
are drawn with dotted curves, with the unmarked areas inside
corresponding to coexistence regions of the two phases at either
side.   The dashed curves are not phase transitions, but disorder
lines between the near-half-filled disordered and the hole-rich or
electron-rich disordered phases. Noteworthy is the narrowness of the
first-order transitions, with jumps in the electron density of the
order of a few percent ({\it i.e.}, the width of the coexistence
region).  The antiferromagnetic phase is unstable to about 8-15\%
hole (or electron) doping away from half-filling. In the
intermediate $U_0/t$ regime, the $\tau_{_{\text{Hb}}}$ phase appears
for about 10-18\% hole (or electron) doping.  At larger $U_0/t$, the
$\tau_{_{\text{tJ}}}$ phase dominates, and exists between 30-35\%
hole (or electron) doping.}

\end{figure*}

\begin{figure}
\centering

\includegraphics*[scale=0.9]{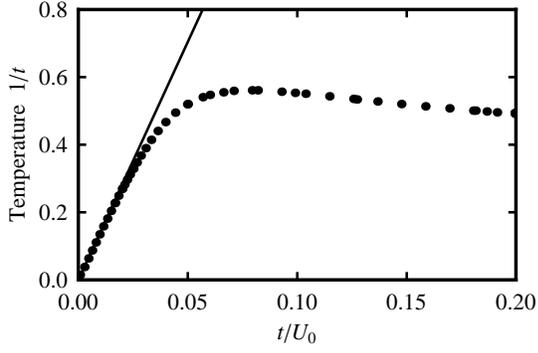}

\caption{The data points are the calculated antiferromagnetic
transition temperatures at half-filling.  The linear relation that
is expected for strong coupling at low temperatures (Sec.V) is
obtained.}
\end{figure}

\begin{figure}
\centering

\includegraphics*[scale=0.9]{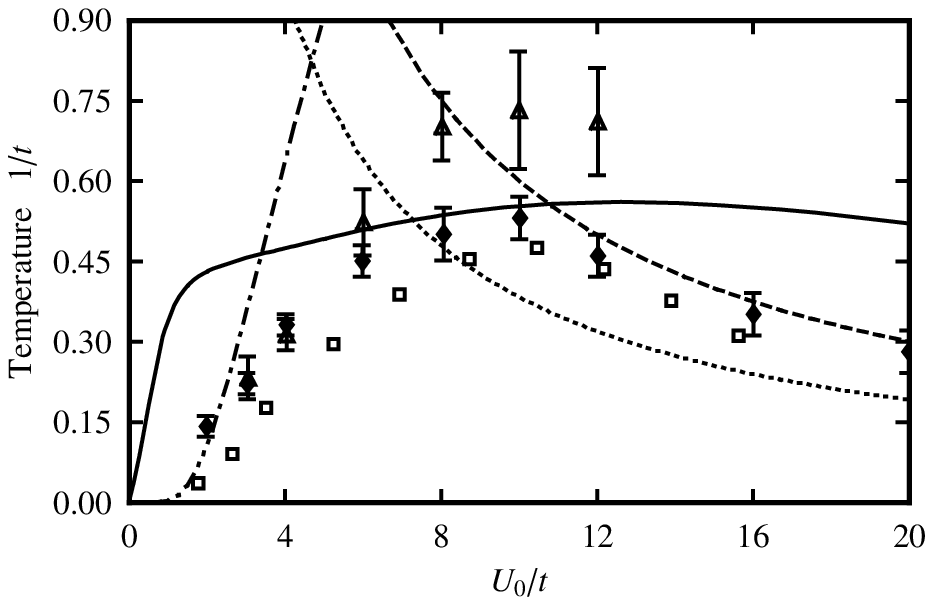}

\caption{Comparison of the antiferromagnetic transition temperatures
at half-filling for the $d=3$ Hubbard model calculated from various
approaches:  the renormalization-group method of the present paper
(solid line); QMC~\cite{Scalettar} (diamonds); QMC~\cite{HirschQMC}
(triangles); DMFT~\cite{Jarrell} (squares); RPA weak-coupling
expansion~\cite{DareAlbinet} (dot-dashed line); and two
approximations for the strong-coupling behavior---the
high-temperature expansion of the Heisenberg model, $1/t =
3.83t/U_0$~\cite{HirschQMC} (dotted line) and Weiss mean-field
theory, $1/t = 6t/U_0$ (dashed line).}
\end{figure}

For $d=3$ and a range of couplings $U_0/t = 5$ to $20$, Figs.~1 show
Hubbard phase diagrams in terms of temperature ($1/t$) versus
chemical potential ($\mu_0/U_0$).  The corresponding phase diagrams
in temperature ($1/t$) versus electron density $\langle n_i \rangle$
are in Fig.~2.  The values of the interaction constants for each
observed phase sink are listed in Table V.  The expectation values
for each phase sink, also listed in Table V, allow us to identify
the phases as follows:

{\bf Hole-rich disordered (hD) phase:}  The electron density $\langle n_i \rangle$ is zero at the
sink and, concomitantly, the electron densities $\langle n_i \rangle$ calculated inside
this phase are low.

{\bf Near-half-filled disordered (nHD) phase:}  The basin of attraction of nHD occurs at
$\mu_0/U_0 \neq 1/2$.  The electron density $\langle n_i \rangle$ is 1 at the sink and,
concomitantly, the electron densities $\langle n_i \rangle$ calculated inside this phase
are closer to half-filling.  {\bf Half-filled disordered (HD) phase:}  The sink is for the disordered phase at perfect
half-filling, $\mu_0/U_0=1/2$ and $\langle n_{i} \rangle = 1$.

{\bf Electron-rich disordered (eD) phase:}  The electron density $\langle n_i \rangle$ is 2
at the sink and, concomitantly, the electron densities $\langle n_i \rangle$ calculated
inside this phase are high.

{\bf Antiferromagnetic (AF) phase:}  The electron density $\langle
n_i\rangle$ is 1 at the sink and, concomitantly, the electron
densities $\langle n_i \rangle$ calculated inside this phase are
closer to half-filling.  The expectation value for the
nearest-neighbor spin-spin correlation is $\langle
\vec{S_i}\cdot\vec{S_j} \rangle = \frac{1}{4}$ at the sink.  Note
that the latter two spins are, on the original cubic lattice,
distant spins on the same sublattice; from this,
antiferromagnetism, $\langle \vec{S_i}\cdot\vec{S_j} \rangle < 0$
when the spins are on different sublattices of the original cubic
lattice, is calculationally obtained throughout this phase.  Since
there is no explicit antiferromagnetic coupling in the initial
Hubbard Hamiltonian, the antiferromagnetic phase is completely a
quantum mechanical effect resulting from the kinetic energy term.
In fact, at half-filling, second-order perturbation theory in $t$,
valid for small $t/U_0$, must yield an effective antiferromagnetic
coupling proportional to $t^2/U_0$.  Thus, for small $t/U_0$,
$t^2/U_0$ should equal the same constant at all antiferromagnetic
phase transitions at half-filling (Recall that all of our coupling
constants are dimensionless, incorporating the inverse temperature
factor $1/kT$).  Equivalently, $t/U_0$ should be linear in $1/t$
at all antiferromagnetic phase transitions at half-filling, for
small $t/U_0$ and therefore for small $1/t$ (low temperature):
\begin{equation}
\label{eq:15b} 1/t\:\sim\:t/U_0\:.
\end{equation}
This is indeed rendered by our calculation, as seen in Fig.~3. For
higher values of $1/t$, Eq.(\ref{eq:15b}) is not applicable, since
second-order perturbation theory does not hold, and indeed our
calculated curve in Fig.~3 deviates from linearity. On the other
hand, the approximation in our recursion relation is even more
justified, since the commutation relations that are ignored involve
terms of order $t^2$.

The antiferromagnetic transition temperature as a function of
coupling $U_0/t$ is also shown in Fig.~4, together with calculated
values from other approximation schemes for the $d=3$ Hubbard model.
We see that our results for intermediate coupling are comparable to
those of quantum Monte Carlo studies~\cite{Scalettar,HirschQMC}. As
expected, our transition temperature vanishes in the limit $U_0/t
\to 0$, since there are no phase transitions for the non-interacting
system.  Thus, our approximation behaves correctly both at strong
coupling (previous paragraph) and at weak coupling.

{\bf $\mathbf\tau_{\text{Hb}}$ and $\mathbf\tau_{\text{tJ}}$
phases:}  For large values of $U_0/t$, the novel phase found in
the $tJ$ model~\cite{FalicovBerker} (which we call
$\mathbf\tau_{\text{tJ}}$) also occurs in the Hubbard model.  In
addition, we find a closely related phase
($\mathbf\tau_{\text{Hb}}$), unique to the Hubbard model, at
smaller $U_0/t$.  The two phases are characterized by very similar
properties:  the hopping strengths $t_0$, $t_2$, and $t_x$
renormalize to $\pm \infty$, and the phase sinks have a non-zero
vacancy hopping expectation value {\small\begin{equation}
\label{eq:16} \left\langle \sum_\sigma h_{i -\sigma}h_{j
-\sigma}(c^\dagger_{i\sigma}
c_{j\sigma}+c^\dagger_{j\sigma}c_{i\sigma})\right\rangle =
\begin{cases} -2/3 & \text{($\mathbf\tau_{\text{tJ}}$)} \\
0.663972 & \text{($\mathbf\tau_{\text{Hb}}$)} \end{cases}\:,
\end{equation}}\noindent for $\mu_0/U_0<1/2$, and a non-zero pair hopping expectation value
{\small\begin{equation} \label{eq:16b} \left\langle \sum_\sigma
n_{i -\sigma}n_{j -\sigma}(c^\dagger_{i\sigma}
c_{j\sigma}+c^\dagger_{j\sigma}c_{i\sigma})\right\rangle =
\begin{cases} 2/3 & \text{($\mathbf\tau_{\text{tJ}}$)} \\
-0.663972 & \text{($\mathbf\tau_{\text{Hb}}$)}
\end{cases}\:,\end{equation}}
\hspace{-1em} for $\mu_0/U_0>1/2$.  In both cases, as expected for
the occurrence of hopping, the electron densities at the sinks have
values different from 0 (empty), 1 (half filled), or 2 (doubly
occupied): $\langle n_i \rangle = 2/3$, $4/3$ and $\langle n_i n_j
\rangle = 1/3$, $5/3$ for the $\mathbf\tau_{\text{tJ}}$ phase, and
$\langle n_i \rangle = 0.668014$, $1.331986$ and $\langle n_i n_j
\rangle = 0.336028$, $1.663972$ for the $\mathbf\tau_{\text{Hb}}$
phase. (At the sinks of non-$\tau$ phases, the electron density is,
on the other hand, 0, 1, or 2.) There are also small spin
correlations in the phase sink limits, $\langle\vec{S}_i
\cdot\vec{S}_j \rangle = -1/4$ for $\mathbf\tau_{\text{tJ}}$, and
$\langle\vec{S}_i \cdot\vec{S}_j \rangle = 0.0840069$ for
$\mathbf\tau_{\text{Hb}}$, which yield, throughout these phases,
small antiferromagnetic correlations in the original system.

The boundaries in Fig.~1 are controlled by fourteen unstable fixed
points, given in Table VI.  For smaller values of $U_0/t$, the
topology of the phase diagram is that of Fig.~1(e), where the
AF/HD, AF/nHD, AF/$\tau_{\text{Hb}}$, and hD/$\tau_{\text{Hb}}$
boundaries are respectively controlled by the second-order fixed
points C$_1^\ast$, C$_2^\ast$, C$_3^\ast$, and C$_4^\ast$.  The
latter three boundaries intersect at the multicritical point
B$_2$, controlled by the fixed point B$_2^\ast$.  A segment of the
hD/nHD boundary just above this intersection is second-order,
controlled by the fixed point C$_5^\ast$, ending at the
multicritical point B$_1$, controlled by the fixed point
B$_1^\ast$. The high-temperature section of the hD/nHD boundary is
a disorder line, controlled by the null fixed point N*, {\it
i.e.}, there is no phase transition above B$_1$.

\begingroup
\renewcommand{\arraystretch}{1.2}
\squeezetable
\begin{table*}
\begin{tabular}{|c|*{10}{c|}c|} \hline
 Phase sink & \multicolumn{10}{c|}{Interaction constants} &
 \parbox{0.7in}{Additional properties}\\ \cline{2-11}
 & $t_0$ & $t_1$ & $t_2$ & $t_x$ & $U$ & $\mu$ & $J$ & $V_2$ & $V_3$ & $V_4$ &\\
\hline hole-rich& 0 & 0 & 0 & 0 & $\infty$ & $-\infty$ & 0 & 0 &
0 & 0 &\\
 disordered hD & & & & & & \tiny{$\mu/U \rsp = \text{const.}$} &&&& &\\
\hline near-half-filled& $2\ln 3$ & 0 & $-2\ln 3$ & $\infty$ &
$\infty$ & $\infty$ & 0 & $-\infty$ & $\infty$ &
-$\infty$ & \tiny{$U\<-2\mu\<-2V_2\<-V_3\<=0$}\\
 disordered nHD &  & &  & \tiny{$\approx 0.24 U$} &
& \tiny{$\approx 0.62 U$} & & \tiny{$\approx -0.47 U$} &
\tiny{$\approx 0.69 U$} &
\tiny{$\approx -1.38 U$} & \tiny{$2V_3\<+V_4\<=0$}\\
\hline half-filled & 0& 0 & 0 & 0 & $\infty$ & $\infty$ & 0 & 0 &
0 &
0 & \tiny{$U\<-2\mu\<=0$}\\
disordered HD &  & &  &  & & \tiny{$= \frac{1}{2} U$} & & &
 & & \\
\hline electron-rich& 0 & 0 & 0 & 0 & $\infty$ & $\infty$ & 0 & 0
& 0 & 0 &\\
 disordered eD &  & &  &  & & \tiny{$\mu/U \rsp =\text{const.}$} & & &  & &\\
\hline antiferro-& $-\infty$ & 0 & $\infty$ & $\infty$ & $\infty$
& $\infty$ & $\infty$ & $-\infty$ &
$\infty$ & $-\infty$ & \tiny{$U\<-2\mu\<-2V_2\<-V_3 \<\to 0$}\\
magnetic AF& \tiny{$\approx -0.29 U$} & & \tiny{$\approx 0.29 U$}
& \tiny{$\approx 0.14 U$} & & \tiny{$\approx 0.57 U$} &
\tiny{$\approx 0.29 U$} & \tiny{$\approx
-0.071 U$} & \tiny{$V_3/U \to 0$} & \tiny{$V_4/U \to 0$} & \tiny{$2V_3\<+V_4\<\to 0$}\\
\tiny{$(\mu_0/U_0 \ne 1/2)$}& &&&&&&&&&& \tiny{$t_2\<-t_0\<\to 0$}\\

\hline antiferro-& 0 & 0 & 0 & 0 & $\infty$ & $\infty$ & $\infty$
& $-\infty$ &
$\infty$ & $-\infty$ & \tiny{${U\<-2\mu\<-2V_2\<-V_3 \<= 0}$}\\
magnetic AF&  & & & & & \tiny{$\approx \frac{1}{2} U$} &
\tiny{$J/U \to 0$} & \tiny{$\approx
-\frac{1}{4} J$} & \tiny{$\approx \frac{1}{2}J$} & \tiny{$\approx -J$} & \tiny{$2V_3\<+V_4 \<= 0$}\\
\tiny{$(\mu_0/U_0 = 1/2)$}& &&&&&&&&&& \tiny{$t_2\<-t_0 \<= 0$}\\

\hline $\tau_{{}_{\text{Hb}}}$ & $-\infty$ & 0 & $\infty$ &
$-\infty$ & $\infty$ & $\infty$ & $\infty$ & $-\infty$ &
$-\infty$ & $\infty$ & \tiny{$t_0 \<+ \mu\<+\frac{1}{4}J\<+V_2 $}\\
\tiny{$(\mu_0/U_0<1/2)$} & \tiny{$\approx -\frac{1}{4}U$} & &
\tiny{$\approx \frac{1}{2}U$} & \tiny{$\approx -\frac{1}{2}U$} & &
\tiny{$\approx \frac{1}{4}U$} & \tiny{$\approx \frac{1}{2} U$} &
\tiny{$\approx
-\frac{1}{8} U$} & \tiny{$V_3/U \to 0$} & \tiny{$V_4/U \to 0$} & \tiny{$\qquad \<\approx -4.35$}\\
\hline $\tau_{{}_{\text{Hb}}}$ & $-\infty$ & 0 & $\infty$ &
$-\infty$ & $\infty$ & $\infty$ & $\infty$ & $-\infty$ &
$\infty$ & $\infty$ & \tiny{$-t_2 \<+U \<-\mu\<+\frac{1}{4}J$}\\
\tiny{$(\mu_0/U_0>1/2)$} & \tiny{$\approx -\frac{1}{2}U$} & &
\tiny{$\approx \frac{1}{4}U$} & \tiny{$\approx -\frac{1}{2}U$} & &
\tiny{$\approx U$} & \tiny{$\approx \frac{1}{2} U$} &
\tiny{$\approx
-\frac{1}{8} U$} & \tiny{$V_3/U \to 0$} & \tiny{$V_4/U \to 0$} & \tiny{$\<-V_2\<-V_3\<\approx -4.35$}\\

\hline $\tau_{{}_{\text{tJ}}}$ & $\infty$ & 0 & $-\infty$ &
$\infty$ & $\infty$ & $-\infty$ & $-\infty$ & $-\infty$ &
$-\infty$ & $-\infty$ & \\
\tiny{$(\mu_0/U_0 < 1/2)$} & \tiny{$\approx 0.13 U$} & &
\tiny{$\approx -1.46 U$} & \tiny{$\approx 0.52 U$} & &
\tiny{$\approx -0.022 U$} & \tiny{$\approx -0.87 U$} &
\tiny{$\approx -0.50 U$} & \tiny{$\approx -1.13 U$} & \tiny{$\approx -0.21 U$} & \\

\hline $\tau_{{}_{\text{tJ}}}$ & $\infty$ & 0 & $-\infty$ &
$\infty$ & $\infty$ & $\infty$ & $-\infty$ & $-\infty$ &
$\infty$ & $-\infty$ & \\
\tiny{$(\mu_0/U_0 > 1/2)$} & \tiny{$\approx 0.42 U$} & &
\tiny{$\approx -0.038 U$} & \tiny{$\approx 0.15 U$} & &
\tiny{$\approx 1.62 U$} & \tiny{$\approx -0.25 U$} &
\tiny{$\approx -0.86 U$} & \tiny{$\approx 0.39 U$} & \tiny{$\approx -0.060 U$} & \\

\hline \multicolumn{12}{c}{}\\[-5pt]
\cline{1-11} Phase sink &
\multicolumn{10}{c|}{Expectation values}\\
\cline{2-11} & $\langle T_0 \rangle$ & $\langle T_1 \rangle$ &
$\langle T_2 \rangle$ & $\langle T_x \rangle$& $\langle
n_{i\uparrow}n_{i\downarrow}\rangle$& $\langle n_i\rangle$ &
$\langle \vec{S}_i\cdot\vec{S}_j \rangle$ & $\langle n_i n_j
\rangle$ & $\langle n_{i\uparrow}n_{i\downarrow}n_j\rangle$ &
$\langle n_{i\uparrow}n_{i\downarrow}n_{j\uparrow}n_{j\downarrow} \rangle$\\
\cline{1-11} hD & 0 & 0 & 0 & 0 & 0 & 0 & 0 & 0 & 0 & 0\\
 nHD & 0 & 0 & 0 & 0 & 0 & 1 & 0 & 1 & 0 & 0\\
 HD & 0 & 0 & 0 & 0 & 0 & 1 & 0 & 1 & 0 & 0\\
 eD & 0 & 0 & 0 & 0 & 1 & 2 & 0 & 4 & 2 & 1\\
 AF  & 0 & 0 & 0 & 0 & 0 & 1 & $\frac{1}{4}$ & 1 & 0 &
0\\
  $\tau_{_{\text{Hb}}}$ & $\left\{ \begin{array}{ll} \parbox{0.2in}{\centering $0.663$\\$\;\;\;972$}\\ 0\\ \end{array}\right.$
 & 0 & $\left\{ \begin{array}{ll} 0\\ \parbox{0.2in}{\centering $-0.66$\\$\;\;3972$}\\ \end{array}\right.$
 & 0 & $\left\{ \begin{array}{ll} 0\\ \parbox{0.2in}{\centering $0.331$\\$\;\;\;986$}\\ \end{array}\right.$ &
$\left\{ \begin{array}{ll} \parbox{0.2in}{\centering $0.668$\\$\;\;\;014$}\\[5pt] \parbox{0.2in}{\centering $1.331$\\$\;\;\;986$}\\
\end{array}\right.$ &
 \parbox{0.25in}{\centering $0.0840$\\$\;\;\;069$} &
  $\left\{ \begin{array}{ll} \parbox{0.2in}{\centering $0.336$\\$\;\;\;028$}\\[5pt]
  \parbox{0.2in}{\centering $1.663$\\$\;\;\;972$}\\ \end{array}\right.$ &
  $\left\{ \begin{array}{ll} 0\\ \parbox{0.2in}{\centering $0.331$\\$\;\;\;986$}\\
  \end{array}\right.$ & 0\\
 $\tau_{{}_{\text{tJ}}}$ & $\left\{ \begin{array}{ll} -\frac{2}{3}\\ 0\\ \end{array}\right.$ & 0 & $\left\{ \begin{array}{ll} 0\\ \frac{2}{3}\\ \end{array}\right.$ & 0 & $\left\{ \begin{array}{ll} 0\\ \frac{1}{3}\\ \end{array}\right.$ &
 $\left\{ \begin{array}{ll} \frac{2}{3}\\ \frac{4}{3}\\ \end{array}\right.$ & $-\frac{1}{4}$ & $\left\{ \begin{array}{ll} \frac{1}{3}\\ \frac{5}{3}\\ \end{array}\right.$ & $\left\{ \begin{array}{ll} 0\\ \frac{1}{3}\\ \end{array}\right.$ & 0\\
%% $\tau_{{}_{{\text{tJ}}_0}}$ & $-\frac{2}{3}$ & 0 & 0 & 0 & 0 &
%%$\frac{2}{3}$ & $-\frac{1}{4}$ & $\frac{1}{3}$ & 0 & 0\\
\cline{1-11}
\end{tabular}
\caption{Interaction constants and expectation values at the phase
sink fixed points.  For $\tau_{_{\text{Hb}}}$ and
$\tau_{{}_{\text{tJ}}}$, the values for $\mu_0 / U_0 {< \atop >}
\frac{1}{2}$ are given.  The hopping expectation values $\langle
T_\alpha \rangle$ are: $\langle T_0 \rangle = \sum_\sigma \langle
h_{i -\sigma}h_{j -\sigma}(c^\dagger_{i\sigma}
c_{j\sigma}+c^\dagger_{j\sigma}c_{i\sigma})\rangle$, $\langle T_1
\rangle = \sum_\sigma \langle
(n_{i\,-\sigma}h_{j\,-\sigma}+h_{i\,-\sigma}n_{j\,-\sigma})
(c^\dagger_{i\sigma}
c_{j\sigma}+c^\dagger_{j\sigma}c_{i\sigma})\rangle$, $\langle T_2
\rangle = \sum_\sigma \langle
n_{i\,-\sigma}n_{j\,-\sigma}(c^\dagger_{i\sigma}
c_{j\sigma}+c^\dagger_{j\sigma}c_{i\sigma})\rangle$, $\langle T_x
\rangle = \langle c^\dagger_{i\uparrow}
c_{j\uparrow}c^\dagger_{i\downarrow}
c_{j\downarrow}+c^\dagger_{j\uparrow}
c_{i\uparrow}c^\dagger_{j\downarrow} c_{i\downarrow} \rangle$.}
\end{table*}
\endgroup

\begingroup
\renewcommand{\arraystretch}{1.2}
\squeezetable
\begin{table*}
\begin{tabular}{|c|c|c|*{10}{c|}c|c|}
\hline & Basin & Type & \multicolumn{10}{c|}{Interaction
constants} &
\parbox{0.8in}{\centering Additional properties}& \parbox{0.55in}{\centering Relevant eigenvalue exponents (y)}\\
 \cline{4-13}
& & & $t_0$ & $t_1$ & $t_2$ & $t_x$ & $U$ & $\mu$ & $J$ & $V_2$ & $V_3$ & $V_4$ & &\\
\hline $F_1^\ast$ & portion of & 1st & $2 \ln 3$ & 0 & $-2 \ln 3$
& $\infty$ & $\infty$ & $-\infty$ & 0 & $\infty$ &
$-\infty$ & $-\infty$ & \tiny{$2\mu\<+V_2 \<\approx -0.396$} & 3\\
& hD/nHD & order & & & & & & &&&
& & \tiny{$U\<-2\mu\<-2V_2\<-V_3 \<\to 0$} &\\
& boundary & &&&&&&&&&&&&\\
\hline $F_2^\ast$ & AF/hD & 1st & $-\infty$ & 0 & $\infty$ &
$\infty$ & $\infty$ & $-\infty$ & $\infty$ & $\infty$ &
$-\infty$ & $\infty$ & \tiny{$U\<-2\mu\<-2V_2\<-V_3 \<\to 0$}  & 3\\
& boundary & order & &  &  & & & &&& & & \tiny{$8\mu\<+ J \<+ 4V_2 \<\approx -0.658$} & \\
\hline $C_1^\ast$ & AF/HD & 2nd & 0 & 0 & 0 & 0 & $\infty$ &
$\infty$ & 1.376 & $-0.0650$ & $0.130$ & $-0.260$ &
\tiny{$U\<-2\mu\<-2V_2\<-V_3\<=0$} & 0.715 \\ & boundary & order & &
& & & &$\approx \frac{1}{2}U$& & &
& & \tiny{$2V_3\<+V_4\<=0$} &\\
\hline $C_2^\ast$ & AF/nHD & 2nd & $-0.554$ & 0 & 0.554 & $\infty$
& $\infty$ & $\infty$ & 1.376 & $-\infty$ &
$\infty$ & $-\infty$ &  \tiny{$U\<-2\mu\<-2V_2\<-V_3 \<\to 0$} & 0.715 \\
&  boundary & order & & & &  & & & & & & & \tiny{$2V_3\<+V_4 \<\to 0$} &\\
\hline $C_3^\ast$ & AF/$\tau_{_{\text{Hb}}}$ & 2nd & $-\infty$ & 0
& $\infty$ & $-\infty$ & $\infty$ & $\infty$ & $\infty$ &
$-\infty$ &
$\infty$ & $-\infty$ & \tiny{$t_0\<+\mu\<+\frac{1}{4}J \<+ V_2 \<\approx -0.739$} & 1.68\\
& boundary & order &  & & &  & & & & & &
 & &\\
\hline $C_4^\ast$ & hD/$\tau_{_{\text{Hb}}}$ & 2nd & $-\infty$ & 0
& $\infty$ & $-\infty$ & $\infty$ & $-\infty$ & $\infty$ &
$\infty$ &
$-\infty$ & $\infty$ & \tiny{$t_0\<+\mu\<+\frac{1}{4}J \<+ V_2 \<\approx -5.178$} & 1.42\\
& boundary & order & & & &  & & & & & & & \tiny{$8\mu \<+ J \<+4V_2 \<\approx 6.617$} &\\
\hline $C_5^\ast$ & portion of & 2nd & $-1.610$ & 0 & $-2\ln 3$ &
$-1.594$ & $\infty$ & 0.523 & 0 & 0.0108 &
$-0.569$ & $-\infty$ & & 1.56\\
& hD/nHD & order & & &  & & & &&& &
\tiny{$\approx -3 U$} & &\\
& boundary & &&&&&&&&&&&&\\

\hline $C_6^\ast$ & hD/$\tau_{_{\text{tJ}}}$ & 2nd & $2.959$ & 0 &
$-29.585$ & $9.629$ & $\infty$ & $1.016$ & $-13.692$ & $-8.332$ &
$-18.259$ & $-\infty$ & & 1.01\\
& boundary & order &  & & &  & & & & & &
 & &\\

\hline $N^\ast$ & portion of & null & 0 & 0 & $-2\ln 3$ & 0 &
$\infty$ & 0 & 0 & 0 &
$4 \ln\frac{\sqrt{3}}{2}$ & $-\infty$ & & 2\\
& hD/nHD& & & &  & & & &&&  &
\tiny{$\approx -3 U$} & &\\
& boundary & &&&&&&&&&&&&\\

\hline $L^\ast$ & $F_1^\ast$, $F_2^\ast$, $C_2^\ast$ & critical &
$-0.554$ & 0 & 0.554 & $\infty$ & $\infty$ & $-\infty$ & 1.376 &
$\infty$ &
$-\infty$ & $-\infty$ &  \tiny{$U\<-2\mu\<-2V_2\<-V_3 \<\to 0$} & 3\\
&  basins meet & endpoint & & & &  & & & & & & & \tiny{$8\mu\<+J\<+4V_2 \<\approx -0.798$} & 0.715155 \\

\hline $B_1^\ast$ & $C_5^\ast$, $N^\ast$ & multi- & $-1.236$ & 0 &
$-2 \ln 3$ & $-1.005$ & $\infty$ & 0.221 & 0 &
0.127 & $-0.652$ & $-\infty$ & & 1.73\\
& basins meet & critical & & & & & & &&&
&\tiny{$\approx -3 U$} & & 0.22\\
\hline $B_2^\ast$ & $C_2^\ast$, $C_3^\ast$, $C_4^\ast$, & multi- &
$-2.156$ & 0 & $-1.555$ & $-2.708$ & $\infty$ & $1.559$ & $0.321$
& $-0.762$ &
$0.201$ & $-\infty$ & & 1.15\\
& $C_5^\ast$ basins meet & critical & & & & & & &&& &
\tiny{$\approx -3 U$} & & 0.27\\

\hline $B_3^\ast$ & $F_1^\ast$, $N^\ast$ & multi- & 0 & 0 & $-2
\ln 3$ & 0 & $\infty$ & $-0.681$ & 0 &
$1.089$ & $-0.438$ & $-\infty$ & \tiny{$3U\<+V_4\<\approx 3.044$} & 2.56\\
& basins meet & critical & & & & & & &&&
&\tiny{$\approx -3 U$} & & 0.96\\

\hline $B_4^\ast$ & $F_2^\ast$, $C_3^\ast$, $C_4^\ast$ & multi- &
$-\infty$ & 0 & $\infty$ & $-\infty$ & $\infty$ & $-\infty$ &
$\infty$ &
$\infty$ & $-\infty$ & $-\infty$ & \tiny{$t_0\<+\mu\<+\frac{1}{4}J\<+V_2 \<\to 0$} & 2.68\\
& basins meet & critical & & && & & &&&
& & \tiny{$8\mu\<+J\<+4V_2 \<\to 0$} & 1.90\\

\hline
\end{tabular}
\caption{Unstable fixed points.  The fixed points of the $\mu_0/U_0
\le 1/2$ half space are given here.}
\end{table*}
\endgroup

\renewcommand{\arraystretch}{1}

As $U_0/t$ is increased, the phase diagram topology becomes more
complex. For $U_0/t \gtrsim 6$, the $\tau_{\text{tJ}}$ phase
appears, its boundary with hD controlled by the second-order fixed
point C$_6^\ast$.  Portions of the lower-temperature boundary
between the hD and nHD phases become first-order (fixed point
F$_1^\ast$), and islands of AF appear above the $\tau_{\text{tJ}}$
phase; their boundaries with hD are also first-order (fixed point
F$_2^\ast$). The intersections of these first-order boundaries with
other phase boundaries are controlled by the additional
multicritical points $B_3^\ast$ and $B_4^\ast$, and by the critical
endpoint $L^\ast$~\cite{BerkerWortis}.

As the coupling $U_0/t$ varies, a most interesting aspect of the
changing phase diagram topology is the relative sizes of the
$\tau_{\text{hB}}$ and $\tau_{\text{tJ}}$ phases.  The
$\tau_{\text{hB}}$ phase is largest at intermediate values of
$U_0/t$, and gradually decreases in size as we move into the
strong-coupling regime, breaking up into narrow slivers until at
large values of $U_0/t$ only tiny remnants of it are left in the
phase diagram.  The $\tau_{\text{tJ}}$ phase appears at intermediate
values of $U_0/t$, grows in size as $U_0/t$ is increased, and
occupies a prominent place in the diagram next to the AF phase in
the strong-coupling regime.  As discussed in Section VII, this is
precisely what we expect, since the Hubbard phase diagram should
approximately reproduce the $tJ$ model results~\cite{FalicovBerker}
in the large $U_0/t$ limit.

Phase diagrams in terms of temperature versus electron density
$\langle n_i \rangle$ are shown in Fig.~2.  It is seen that the
antiferromagnetic phase is unstable to at most 15\% hole (or
electron) doping at low temperatures.  The $\tau_{_{\text{Hb}}}$ and
$\tau_{_{\text{tJ}}}$ phases exist at different doping values, with
$\tau_{_{\text{Hb}}}$ appearing for approximately 10-18\% doping,
directly adjacent to the AF phase, and $\tau_{_{\text{tJ}}}$ in the
30-35\% doping range.  The narrowness of the first-order
transitions, with jumps in the electron density of the order of a
few percent, is noteworthy.

\section{Specific Heat Results}

From the calculated expectation values of the operators occurring in
the Hamiltonian [Eq.(\ref{eq:2})], we have obtained the
dimensionless internal energy per bond $\langle \beta
H(i,j)\rangle$.  Recall that dimensionless coupling constants are
exhibited in the Hubbard Hamiltonian of Eq.(\ref{eq:2}), {\it e.g.},
\begin{equation}\label{eq:17ex1}
t=\frac{\tilde{t}}{k_B T}\,,
\end{equation}
where $\tilde{t}$ is a constant that does not depend on temperature.
The specific heat is calculated with
\begin{multline}\label{eq:17}
C = \frac{\partial \langle H(i,j)\rangle}{\partial T} =
k_B\left\{\frac{\partial}{\partial t^{-1}} \langle
c^\dagger_{i\sigma}c_{j\sigma}+c^\dagger_{j\sigma}c_{i\sigma}\rangle
\right.\\
+\left.\frac{U}{t} \frac{\partial}{\partial t^{-1}} \langle
n_{i\uparrow}n_{i\downarrow}+n_{j\uparrow} n_{j\downarrow}\rangle
\right\}.
\end{multline}
The partial derivatives are taken at fixed $U_0/t$ and at fixed
density $\langle n_i \rangle$.

In Fig. 5 we plot $\gamma = C/T$ for $U_0/t = 15$ at several
different electron densities.  (The corresponding phase diagram is
shown in Fig. 2(b)).  At half-filling, $\langle n_i \rangle = 1.00$,
we observe a broad peak near the HD/AF transition temperature, which
we can attribute to the onset of spin order. As we dope the system
with holes, this peak gets sharper, becoming most pronounced near
$\langle n_i \rangle = 0.68$, directly above the transition
temperature between the hD and $\tau_{_{\text{tJ}}}$ phases.  In
fact, the $C/T$ curve shows a multipeak structure near the
transition, a general characteristic of the phase diagram region
just above the $\tau_{_{\text{tJ}}}$ phase. At electron density
$\langle n_i \rangle = 0.60$, no longer in the $\tau_{_{\text{tJ}}}$
range, the peak decreases in size and broadens out again.

The distinct nature of the $\tau_{_{\text{tJ}}}$ and
$\tau_{_{\text{Hb}}}$ phases becomes clear when we look at the low
temperature specific heat.  In Fig. 6 we plot the coefficient
$\gamma=C/T$ as a function of electron density for $U_0/t=15$ and at
low temperature $1/t = 0.085$.  In the limit as $T \to 0$, $\gamma$
is a measure of the linear contribution to the specific heat due to
quasiparticle excitations.  Near half-filling, $\gamma$ is close to
zero, increases to a small level with sufficient hole doping, falls
to near zero again in the $\tau_{_{\text{Hb}}}$ phase, and
dramatically increases only after the system makes a narrow
first-order transition to the hole-rich disordered phase.  The
steady rise of $\gamma$ in the hD phase with further hole doping is
consistent with a Fermi liquid interpretation of this phase.  The
increase in $\gamma$ is interrupted by the $\tau_{_{\text{tJ}}}$
interval, where the curve makes a sharp oscillation, but continues
in the hD region on the other side.

\begin{figure}
\centering

\includegraphics*[scale=0.9]{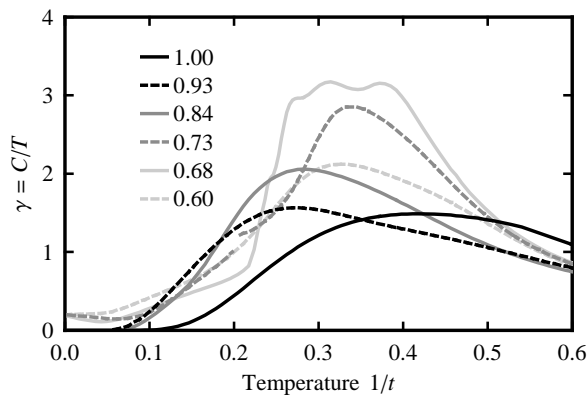}

\caption{The specific heat coefficient $\gamma = C / T$ as a
function of temperature for $U_0/t = 15$, at several different
electron densities $\langle n_i \rangle$ indicated in the legend.
For this temperature range the densities $\langle n_i \rangle =
1.00$ and $0.93$ lie inside the antiferromagnetic (AF) phase, $0.84$
inside the $\tau_{_{\text{Hb}}}$ phase, $0.73$ and $0.60$ inside the
hole-rich disordered (hD) phase, and $0.68$ inside the
$\tau_{_{\text{tJ}}}$ phase.  Here and in the following figures,
$\gamma$ is shown in units of $k_B^2/\tilde{t}$, where $\tilde{t}$
is the temperature-independent constant in Eq.~(\ref{eq:17ex1}).}
\end{figure}

\begin{figure}
\centering

\includegraphics*[scale=0.85]{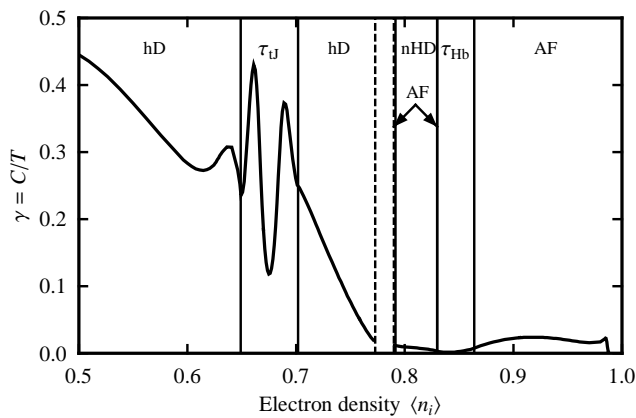}

\caption{The specific heat coefficient $\gamma = C / T$ for
$U_0/t=15$ at the low temperature of $1/t = 0.085$, as a function of
electron density $\langle n_i \rangle$. The corresponding phases are
indicated near the top of the figure, with second-order phase
boundaries marked by thin vertical lines.  The interval between the
vertical dashed lines corresponds to the first-order phase
transition.}
\end{figure}

\begin{figure}
\centering

\includegraphics*[scale=0.9]{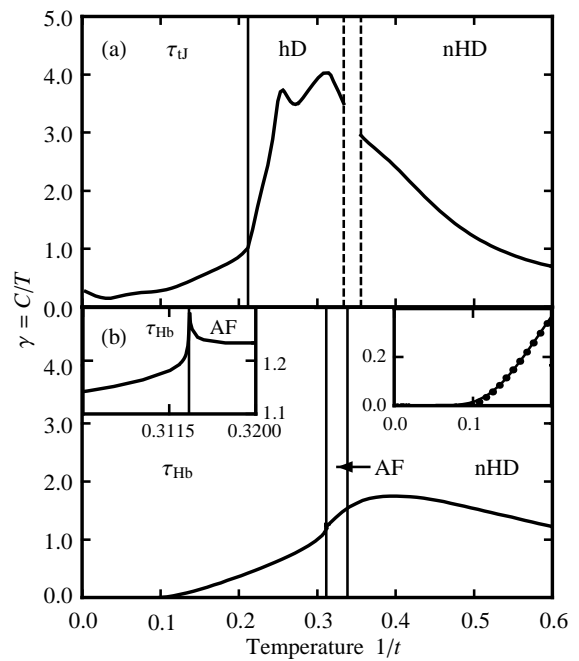}

\caption{The specific heat coefficient $\gamma = C / T$ as a
function of temperature for two different electron densities and
values of $U_0/t$: (a) $\langle n_i \rangle = 0.68$, $U_0/t = 20$;
(b) $\langle n_i \rangle = 0.875$, $U_0/t = 7.5$.  Phases are
indicated near the top of the figures, with second-order phase
boundaries marked by thin vertical lines.  The interval between the
vertical dashed lines corresponds to a first-order phase transition.
In diagram (b) the top left inset shows a close-up of the cusp in
$\gamma$ at the AF/$\tau_{_{\text{Hb}}}$ transition temperature. The
data points in the top right inset are calculated $\gamma$ values
for temperatures $1/t < 0.2$, fitted to a BCS-like exponential curve
of the form $C/k_B T = \frac{A}{T^{5/2}}
\exp\left(-\frac{\Delta}{T}\right)$, with best-fit parameters $A =
1.02 \pm 0.06$ and $\Delta = 1.01 \pm 0.01$, where $t^{-1}$ is used
as the temperature variable.}
\end{figure}

We see that the $\tau_{_{\text{tJ}}}$ phase has non-zero $\gamma$ at
low temperatures, while the $\tau_{_{\text{Hb}}}$ phase does not. In
Figs.~7(a) and (b) we contrast the two $\tau$ phases directly,
comparing representative $C/T$ curves for $\tau_{_{\text{tJ}}}$ and
$\tau_{_{\text{Hb}}}$ transitions.  We observe that in the
$\tau_{_{\text{Hb}}}$ phase the low-temperature specific heat
exhibits an exponential form characteristic of a gap in the
quasiparticle spectrum.  Specific heat data points for temperatures
$1/t < 0.2$, shown in the top right inset of Fig.~7(b), were found
to fit a theoretical curve of the same form as in the $T \to 0$
limit of a weakly-coupled, BCS-type superconductor,
\begin{equation}\label{eq:19}
\frac{C}{k_B} = \frac{A}{T^{3/2}}
\exp\left(-\frac{\Delta}{T}\right)\,,
\end{equation}
with a best-fit coefficient $A = 1.02 \pm 0.06$ and a
zero-temperature gap $\Delta = 1.01 \pm 0.01$, where $t^{-1}$ is
used as the temperature variable. In contrast, the
$\tau_{_{\text{tJ}}}$ phase clearly has a gapless spectrum, as we
see in the $C/T$ curve of Fig.~7(a). As mentioned earlier, we also
clearly see multiple peaks in the specific heat just above the
hD/$\tau_{_{\text{tJ}}}$ transition temperature.

The $\tau_{_{\text{tJ}}}$ and $\tau_{_{\text{Hb}}}$ phases have
similar properties at the phase sink, most notably a non-zero
hopping amplitude, and thus are both good candidates for
superconductivity.  Since the two phases are dominant in different
$U_0/t$ regimes, their contrasting specific heat characteristics can
potentially be understood as the difference between strongly-coupled
and weakly-coupled superconducting phases.  For the
strongly-coupled, BEC-like case, pairing occurs above $T_c$, and
these tightly bound bosonic pairs condense at the transition
temperature.  The double-peak structure in the specific heat above
the $\tau_{_{\text{tJ}}}$ phase is a possible indicator of such pair
formation.  Additionally, we expect that a BEC-like superconducting
transition in three dimensions should have a specific heat critical
exponent $\alpha = -1$~\cite{Junod}. Analysis of the $C_6^\ast$
fixed point, governing the hD/$\tau_{_{\text{tJ}}}$ boundary, yields
the result $\alpha = -0.97$.  The presence of low-lying excitations
in a Bose gas is also consistent with the fact that we do not see a
gap in the low-temperature specific heat of the
$\tau_{_{\text{tJ}}}$ phase.

Turning now to the $\tau_{_{\text{Hb}}}$ phase, we already noted
that its specific heat can be closely fitted at low temperatures to
a BCS-like exponential curve, which is exactly what we would expect
for a weakly-coupled superconducting phase.  Analysis of the
$C_3^\ast$ fixed point, controlling the AF/$\tau_{_{\text{Hb}}}$
boundary, yields a specific heat coefficient $\alpha = -0.27$. This
translates into a finite cusp at the transition temperature, as
shown in the top left inset of Fig.~7(b).  For weak and intermediate
couplings the superconducting transition is expected to belong to
the universality class of the $d=3$ $XY$ model, with $\alpha =
-0.013$ \cite{Lipa} (examples of transitions in this class include
the superfluid transition of $^4$He, the superconducting transition
in certain high-$T_c$ materials like Y-123, and also in conventional
superconductors, though for the latter the critical region is too
narrow to be observed experimentally)~\cite{Junod}.  Our calculated
$\alpha$ is closer to the $d=3$ $XY$ than to the BEC value,
supporting the weak-coupling interpretation of the
$\tau_{_{\text{Hb}}}$ phase.

\section{The $tJ$ Limit of the Hubbard Model}

\begin{figure}
\centering \includegraphics*[scale=0.9]{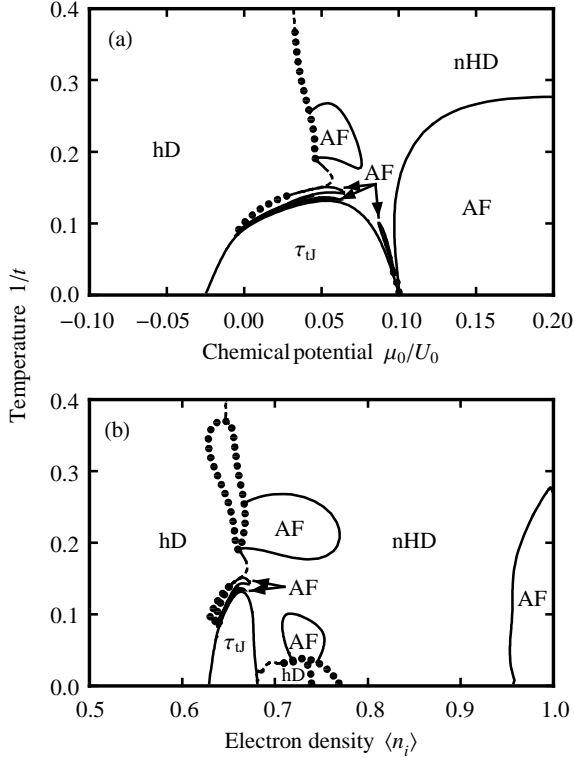}

\caption{$d=3$ Hubbard model phase diagram for large Coulomb
repulsion $U_0/t = 50$ in temperature versus (a) chemical potential,
(b) electron density $\langle n_i \rangle$.  The full curves are
second-order phase boundaries, while the dotted curves indicate
first-order boundaries.  The dashed lines are not phase transitions,
but disorder lines between the near-half-filled disordered and
hole-rich disordered phases.}
\end{figure}

\begin{figure}
\centering \includegraphics*[scale=0.8]{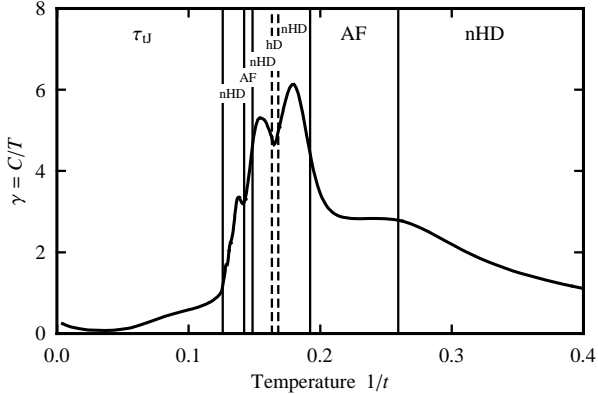}

\caption{The specific heat coefficient $\gamma = C/T$ as a function
of temperature for $U_0/t=50$ and $\langle n_i \rangle = 0.67$.
Phases are indicated near the top of the figure, with second-order
phase boundaries marked by thin vertical lines.  The dashed lines
are not phase transitions, but disorder lines between the
near-half-filled disordered and hole-rich disordered phases.}

\end{figure}

\begin{figure}
\centering
\includegraphics*[scale=0.9]{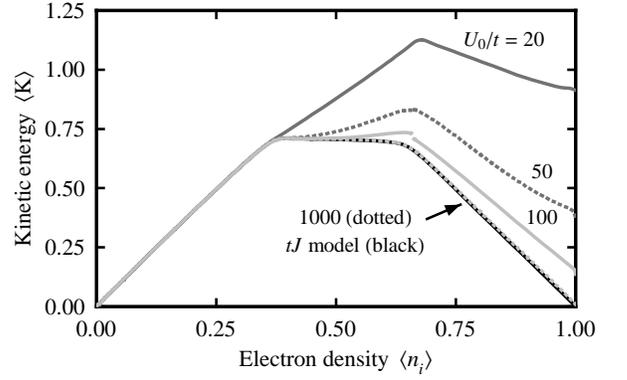}

\caption{The kinetic energy per bond $\langle K \rangle =
-\sum_{\sigma} \langle c_{i\sigma }^{\dagger }c_{j\sigma
}+c_{j\sigma }^{\dagger }c_{i\sigma }\rangle$ as a function of
electron density $\langle n_i \rangle$ at temperature $1/t = 0.2$
for Coulomb repulsions $U_0/t = 20$, 50, 100, and 1000 (indicated by
numbers next to each curve). The solid curve at the bottom is the
result calculated using the $tJ$ model renormalization-group
equations~\cite{FalicovBerker,FalicovBerkerT} at the same
temperature, with the corresponding $J/t = 0.004$.}
\end{figure}

In the strong-coupling limit $U_0 \gg t$, second-order perturbation
theory in $t/U_0$ applied to the Hubbard model leads to the
following Hamiltonian (known as the $tJ$
model)~\cite{Anderson,BZA,FalicovBerker,FalicovBerkerT,Chao,Hirsch2,LMFalicov,RabeBhatt,Bares},
\begin{multline}\label{eq:20}
H_{\text{$tJ$}} = -t\sum_{\langle ij\rangle ,\sigma } P\left(
c_{i\sigma }^{\dagger }c_{j\sigma }+c_{j\sigma }^{\dagger
}c_{i\sigma }\right)P\\
+ J \sum_{\langle ij\rangle}(\vec{S}_i\cdot
\vec{S}_j - \frac{1}{4}n_i n_j)\,,
\end{multline}
where $J = 4t^2/U$ and $P$ is a projection operator prohibiting
double occupation of a lattice site.  In addition to the terms shown
above, the perturbation theory generates a three-site term of the
form $\sum_{\langle i k j \rangle} c^\dagger_{i\sigma}
(S_k)_{\sigma\sigma^\prime} c_{j\sigma^\prime}$, but this term is
usually ignored, from the assumption that it does not radically
alter the physics of the $tJ$ model.  (Our current results, directly
from the strong-coupling limit of the actual Hubbard model, confirm
this assumption.)  We thus expect that our Hubbard model approach in
the limit of large $U_0/t$ should give results qualitatively similar
to those found for the $tJ$ model in earlier renormalization-group
studies~\cite{FalicovBerker,FalicovBerkerT}. The phases of the $tJ$
model found in these studies are identical to those of the Hubbard
model, except that there is no $\tau_{_{\text{Hb}}}$ phase.

Fig.~8 shows the Hubbard model phase diagram in terms of temperature
versus chemical potential and temperature versus electron density
for $U_0/t = 50$. At this large coupling, we do indeed observe a
phase diagram very similar to that found in the earlier study of the
$tJ$ model~\cite{FalicovBerker,FalicovBerkerT}.  In particular, the
$\tau_{\text{\text{tJ}}}$ phase is surrounded by AF islands, and
directly above $\tau_{\text{\text{tJ}}}$ we get a lamellar structure
of alternating AF, nHD, and hD phases.  The AF phase near
half-filling is unstable to only about 5\% hole doping.  This phase
diagram can be seen as an evolution from the $U_0/t=20$ result of
Figs.~1(a) and 2(a), with the $\tau_{\text{\text{hB}}}$ entirely
disappearing at $U_0/t=50$ except for infinitesimal slivers. The
multiple peaks in the specific heat above the
$\tau_{\text{\text{tJ}}}$ transition persist in the strong-coupling
limit, as seen in Fig.~9, which plots the specific heat coefficient
$\gamma$ for $U/t_0=50$ at $\langle n_i \rangle = 0.67$. The peak
structure here is more complex than in Fig.~7(a), due to the
above-mentioned lamellar phases.

We can also observe the evolution from the Hubbard to the $tJ$
limits through the expectation value of the kinetic energy per bond,
$\langle K \rangle = -\sum_{\sigma} \langle c_{i\sigma }^{\dagger
}c_{j\sigma }+c_{j\sigma }^{\dagger }c_{i\sigma }\rangle$, which is
proportional to the density of free carriers in the system.  Fig.~10
shows $\langle K \rangle$ as a function of electron density for the
temperature $1/t = 0.2$, calculated at several different couplings
$U_0/t$. As $U_0/t$ is increased, the value of $\langle K \rangle$
at half-filling is reduced, and when $U_0/t = 1000$ we are close to
the $tJ$ limit, with the kinetic energy at half-filling almost zero,
indicating no available free carriers due to the prohibitively high
energy of double occupation. The $U_0/t = 1000$ curve almost exactly
overlaps the result calculated from the $tJ$ model
renormalization-group equations at the same temperature using the
corresponding coupling $J/t = 4t/U_0 = 0.004$.

\begin{acknowledgments}
This research was supported by the U.S. Department of Energy under
Grant No. DE-FG02-92ER-45473, by the Scientific and Technical
Research Council of Turkey (T\"UBITAK) and by the Academy of
Sciences of Turkey.  MH gratefully acknowledges the hospitality of
the Feza G\"ursey Research Institute and of the Physics Department
of Istanbul Technical University.
\end{acknowledgments}

\appendix

\section*{Appendix A:  Determination of the $\gamma_p$ \hspace{0.2in} in Terms of the Matrix Elements \hspace{0.25in} of the Three-Site
Hamiltonian}

Eq.(8) allows us to express the matrix elements $\gamma_p \equiv
\langle \phi_p | e^{-\beta^\prime H^\prime (i,k)}|\phi_{p} \rangle$
of the renormalized, exponentiated two-site Hamiltonian in terms of
matrix elements of the unrenormalized, exponentiated three-site
Hamiltonian, as given below.  The $\gamma_p$, in turn, determine the
renormalized interaction constants, in Eq.(10).  In the equations
below, $\langle \psi_q | | \psi_{\bar{q}} \rangle$ denotes $\langle
\psi_q | e^{-\beta H(i,j) -\beta H(j,k)} | \psi_{\bar{q}} \rangle$:

\begin{align*}
\gamma_1 =& \langle \psi_1 | | \psi_1 \rangle \rsp+ 2\langle
\psi_2 |
| \psi_2 \rangle \rsp+ \langle\psi_{9} | | \psi_{9} \rangle,\displaybreak[0]\\[5pt]
\gamma_2 =& \langle \psi_{3} | | \psi_{3} \rangle \rsp+
\frac{1}{2}\langle \psi_{8} | | \psi_{8} \rangle
\rsp+\frac{3}{2}\langle
\psi_{15} | | \psi_{15} \rangle \rsp+\langle \psi_{24} | | \psi_{24} \rangle,\displaybreak[0]\\[5pt]
\gamma_4 =& \langle \psi_{6} | | \psi_{6} \rangle \rsp+
\frac{1}{2}\langle \psi_{11} | | \psi_{11} \rangle
\rsp+\frac{3}{2}\langle
\psi_{20} | | \psi_{20} \rangle \rsp+\langle \psi_{32} | | \psi_{32} \rangle,\displaybreak[0]\\[5pt]
\gamma_{6} =& \langle \psi_{10} | | \psi_{10} \rangle  \rsp+
2\langle \psi_{26} | | \psi_{26} \rangle \rsp+\langle\psi_{44} | | \psi_{44} \rangle,\displaybreak[0]\\[5pt]
\gamma_7 =& \langle \psi_{13} | | \psi_{13} \rangle \rsp+ \langle
\psi_{34} | | \psi_{34} \rangle \rsp+ \langle \psi_{38} | |
\psi_{38}\rangle\rsp+\langle\psi_{48} | | \psi_{48}
\rangle,\displaybreak[0]\\\\
\gamma_8 =& \langle \psi_{12} | |
\psi_{12} \rangle \rsp+2\langle
\psi_{31} | | \psi_{31} \rangle \rsp+  \langle \psi_{47} | | \psi_{47} \rangle,\\[5pt]
\gamma_9 =& \langle \psi_{14} | | \psi_{14} \rangle \rsp+
\frac{2}{3}\langle \psi_{23} | | \psi_{23} \rangle \rsp+
\frac{4}{3}\langle \psi_{39} | | \psi_{39} \rangle \rsp+\langle
\psi_{49} | | \psi_{49} \rangle,\displaybreak[0]\\[5pt]
\gamma_{12} =&  \langle \psi_{25} | | \psi_{25} \rangle
\rsp+\frac{1}{2}\langle \psi_{45} | | \psi_{45} \rangle \rsp+
\frac{3}{2}\langle \psi_{50} | | \psi_{50} \rangle \rsp+\langle
\psi_{59} | |\psi_{59} \rangle ,\displaybreak[0]\\[5pt]
\gamma_{14} =& \langle \psi_{33} | | \psi_{33}
\rangle\rsp+\frac{1}{2}\langle \psi_{46} | | \psi_{46} \rangle
\rsp+\frac{3}{2}\langle \psi_{55} | | \psi_{55} \rangle
\rsp+\langle
\psi_{62} | |\psi_{62} \rangle ,\\[5pt]
%%\end{align*}
%%\begin{align*}
\gamma_{16} =& \langle \psi_{43} | | \psi_{43} \rangle
\rsp+2\langle \psi_{58} | | \psi_{58} \rangle
\rsp+\langle\psi_{64} | |
\psi_{64}\rangle,\\[5pt]
\gamma_{0} \equiv& \langle \phi_{6} | e^{-\beta^\prime H^\prime
(i,k)}|\phi_{8}\rangle \\[5pt]
=&\langle \psi_{10} | | \psi_{12} \rangle \rsp+ 2\langle \psi_{26}
| | \psi_{31} \rangle\rsp+\langle \psi_{44} | | \psi_{47} \rangle.
\end{align*}

\end{document}